\documentclass[fleqn,usenatbib]{mnras}

\usepackage{newtxtext,newtxmath}
\usepackage[T1]{fontenc}
\usepackage{ae,aecompl}

\usepackage{graphicx}	% Including figure files
\usepackage{amsmath}	% Advanced maths commands
\usepackage{amssymb}	% Extra maths symbols
\usepackage{longtable}
\usepackage{multicol}

\newcommand{\plus}{\scalebox{0.8}{+}}

\usepackage{txfonts}
\usepackage{xcolor}
\usepackage{soul}

\usepackage[export]{adjustbox}
\title[The 2BIGB $\gamma$-ray catalogue]{Extreme \& High Synchrotron Peak Blazars beyond 4FGL: The 2BIGB $\rm \gamma$-ray catalogue}

 \author[B. Arsioli et al.]{
B. Arsioli$^{1,2}$\thanks{E-mail: arsioli@ifi.unicamp.br, bruno.arsioli@gmail.com},
Y-L. Chang$^{2,3}$%\thanks{E-mail: yuling.chang@ssdc.asi.it},
B. Musiimenta$^{4}$ %\thanks{E-mail: mblessing78@gmail.com},
\\
$^{1}$Instituto de F\'isica Gleb Wataghin, Universidade Estadual de Campinas UNICAMP, Rua S\'ergio Buarque de Holanda 777, Campinas, Brazil \\
$^{2}$ICRANet, P.zza della Repubblica 10, I-65122, Pescara, Italy \\
$^{3}$Tsung-Dao Lee Institute, Shanghai Jiao Tong University, 800 Dongchuan RD. Minhang District, 200240, Shanghai, China \\
$^{4}$Mbarara University of Science and Technology, Department of Physics, P.O. box 1410, Mbarara, Uganda 
}

\date{Accepted 2020 January 31. Received 2019 November 19; in original form }

\pubyear{2020}

\begin{document}
\label{firstpage}
\pagerange{\pageref{firstpage}--\pageref{lastpage}}
\maketitle

% 2BIGB Total: 1160 sources; 925 in 4FGL + 235 new  (9 were reported in previous FGL/FHL cats)
\begin{abstract}
This paper presents the results of a $\rm \gamma$-ray likelihood analysis over all the extreme and high synchrotron peak blazars (EHSP \& HSP) from the 3HSP catalogue. We investigate 2013 multifrequency positions under the eyes of Fermi Large Area Telescope, considering 11 years of observations in the energy range between 500\,MeV to 500\,GeV, which results in 1160 $\rm \gamma$-ray signatures detected down to the TS\,=\,9 threshold. The detections include 235 additional sources concerning the Fermi Large Area Telescope Fourth Source Catalog (4FGL), all confirmed via high-energy TS maps, and represent an improvement of $\sim$25\% for the number of EHSP \& HSP currently described in $\rm \gamma$-rays. We build the $\rm \gamma$-ray spectral energy distribution for all the 1160 2BIGB sources, plot the corresponding $\rm \gamma$-ray logN-logS, and measure their total contribution to the extragalactic gamma-ray background, which reaches up to $\sim$33\% at 100\,GeV. Also, we show that the $\rm \gamma$-ray detectability improves according to the synchrotron peak flux as represented by the Figure of Merit (FOM) parameter, and note that the search for TeV peaked blazars may benefit from considering HSP and EHSP as a whole, instead of EHSPs only. The 2BIGB acronym stands for `Second Brazil-ICRANet Gamma-ray Blazars' catalogue, and all the broadband models and spectral energy distribution data-points will be available on public data repositories (OpenUniverse, GitHub, and Brazilian Science Data Center-BSDC). 
%This is a simple template for authors to write new MNRAS papers.The abstract should briefly describe the aims, methods, and main results of the paper.It should be a single paragraph not more than 250 words (200 words for Letters).No references should appear in the abstract.
\end{abstract}

%From those, 226 are new detections with no counterpart in previous {\it Fermi}-LAT catalogues 1-2-3FGL and the 1-2-3FHL.
% http://www.openuniverse.asi.it/ ; http://bsdc.icranet.org/  ;  https://vizier.u-strasbg.fr/viz-bin/VizieR

%[Revising Abstract]
%We build the $\rm \gamma$-ray spectral energy distribution for all the 1160 sources, delivering to the community a dedicated spectral description of the 3HSP sources. We plot the $\rm \gamma$-ray logN-logS for the 3HSP catalogue, and also for the EHSP \& HSP subsamples. We measure the total contribution of HSP+EHSP to the extragalactic gamma-ray background, which reaches up to $\sim$33\% at 100\,GeV. Also, we show how does the $\rm \gamma$-ray detectability improves according to the synchrotron peak flux as represented by the Figure of Merit (FOM) parameter. We highlight that the search for TeV peaked blazars may benefit from considering HSP and EHSP as a whole, instead of EHSPs only. We entitle this $\rm \gamma$-ray sample as Second Brazil-ICRANet Gamma-ray Blazar catalogue, with the acronym 2BIGB. All information will be available on public data repositories (Brazilian Science Data Center-BSDC, OpenUniverse, GitHub), including the broadband models and the spectral energy distribution data points.


%%%%%%%%%%%%%%%%%%%%%%%%%%%%%%%%%%%%%%%%%%%%%%%%%%

%%%%%%%%%%%%%%%%% BODY OF PAPER %%%%%%%%%%%%%%%%%

\section{Introduction}
\label{section:first}

% \url{https://github.com/BrunoArsioli}

The Fermi Large Area Telescope \citep[LAT, ][]{FermiLAT} produced our most detailed picture of the gamma-ray sky and has opened a window to investigate high \& very-high energy\footnote{We use high-energy (HE) for the 0.2-100\,GeV range, and very high-energy (VHE) for E\,>\,100\,GeV.} processes throughout the universe. Since the delivery of its first source catalogue \citep{1FGL,1lac}, the {\it Fermi}-LAT mission has identified blazars as the main population of extragalactic $\rm \gamma$-ray emitters. There are 5066 $\rm \gamma$-ray point sources reported in the latest {\it Fermi}-LAT catalogue \citep[4FGL,][ gll-pcs-v20.fit]{4FGL}, all detected with significance >4$\sigma$ (TS$>$25) covering 8 years of observations, and integrating over the 50\,MeV to 1\,TeV energy band. According to the Fourth catalogue of active galactic nuclei detected with {\it Fermi}-LAT \citep[4LAC,][]{4lac}, there are 3647 4FGL sources out of the galactic plane (at $|b|>10^{\circ}$), from which 79$\%$ are confident counterparts of active galactic nuclei (AGN). Given all those gamma-ray AGNs, 98\% are blazars. At energies larger than 1\,TeV, there is a similar dominance, as follows from the latest version of TeVcat\footnote{TeVcat:\url{http://tevcat.uchicago.edu/} is a regularly updated list of TeV detected astrophysical sources. The TeVcat reports on 81 extragalactic TeV sources, which includes 72 blazars, 3 Starbursts, 1 Globular Cluster, 2 AGNs of unknown type and 3 gamma-ray bursts.}. The TeVcat lists 225 TeV detected sources, and nearly 90\% of the extragalactic ones (72 out of 81) are blazars. 

Blazars are relatively rare objects with nearly 4000 optically confirmed sources -with available optical spectrum- as listed in 5BZcat and 3HSP catalogues \citep{5BZcat-V5,3HSP}. The 5BZcat lists 3651 confirmed blazars of all types, while the 3HSP lists 2013 objects with a focus to high synchrotron peak blazars and blazar-candidates. Within the 3HSP catalogue there are 657 blazars already listed in 5BZcat, 257 are newly confirmed blazars (out of 5BZcat), and 1099 are blazar-candidates (high confident blazars missing optical identification). Blazars are well established as the dominant population of extragalactic $\gamma$-ray sources, but yet a significant fraction lack detection at HE and VHE.% 

According to the unification scheme for active galactic nuclei, blazars are AGNs producing relativist jets that happen to point close to our line of sight \citep{Urry-Padovani1995,giommisimplified,AGN-What-in-Name-Padovani2017}. The jets are launched from the AGN's core and composed of relativistic charged particles trapped within the jet's magnetic field. This configuration is most likely powered by accretion into supermassive black holes \citep{Blazar-Paradigm-Dermer2015,Blazar-SMBH-Petropoulou2016}, and gives rise to non-thermal (synchrotron) spectral emission extending through the entire electromagnetic spectra, from radio up to TeV $\rm \gamma$-rays.

Blazars can be classified according to the frequency associated with their synchrotron peak and called as low, intermediate, high, and extremely-high synchrotron peak sources (respectively: LSP for $ \rm \nu^{syn.}_{peak}$<$\rm 10^{14}$Hz, ISP for $\rm 10^{14}$<$\rm \nu^{syn.}_{peak}$<$\rm 10^{15}$Hz, HSP for $\rm 10^{15}$<$\rm \nu^{syn.}_{peak}$<$\rm 10^{17}$Hz, and EHSP for $\rm \nu^{syn.}_{peak}$>$\rm 10^{17}$Hz). The EHSP and HSP sources have -on average- a hard\footnote{We refer to hard (and soft) photon index meaning $\Gamma$<2.0 (and $\Gamma$>2.0).} photon index $ \rm \langle \Gamma \rangle$=1.85$\pm$0.01 as probed by {\it Fermi}-LAT \citep{1WHSP,3lac} in the MeV to GeV window. For this reason, EHSP and HSP blazars are considered promising for VHE observations with the Cherenkov Telescope Arrays \citep{CTA,CTA-book-science-2019}, and the community dedicates significant attention to characterize their SEDs via multifrequency observation campaigns \citep[][]{Multiwave-Blazar-Campaign2017,multifreq-campaign-IC310-2017,optical-campaign-blazars2017,multifrequec-campaign-NGC1275-2014,multifreq-Mark421-2000ApJ}. 

This work aims to contribute to the description of blazars at HE \& VHE by presenting new $\gamma$-ray detections, describing their SEDs, and studying their population properties. We use the position from blazars and blazar-candidates as multifrequency seeds for a likelihood analysis with the {\it Fermi}-LAT Science Tools, to search for associated $\rm \gamma$-ray signatures. The 3HSP catalogue \citep{3HSP} is used as reference blazar-database, and is currently the most extensive compilation of EHSP and HSP sources. 

Given the spectral characteristics of 3HSP sources, our analysis focuses on a particular blazars subsample, which is more likely to be detectable at the VHE band. In fact, 925 3HSPs sources already have a 4FGL counterpart (we call them 3HSP-4FGL) meaning that the 3HPS source lies within the 95\% error region reported in 4FGL. So by studying the entire 3HSP catalogue under {\it Fermi}-LAT eyes, we can compare results with 4FGL to show the stability of our analysis and the robustness of all new detections. We compute high-energy TS maps for the new sources and confirm that those signatures emerge from a low TS background with points-like TS distribution, avoiding spurious detections with origin related to underestimated background.  
% 2BIGB Total: 1160 ; 925 in 4FGL + 235 new  

The results of this large-scale analysis is the Second Brazil-ICRANet Gamma-ray Blazar catalogue, 2BIGB, which lists 1160 $\rm \gamma$-ray excess signatures (Table \ref{table:2BIGB-10yrs-slim}) with significance down to TS\,=\,9. Our broadband analysis integrates over the 500\,MeV up to 500\,GeV energy range along 11 years of observations with {\it Fermi}-LAT (Aug. 2008 to Aug. 2019), and we compute the $\rm \gamma$-ray SED for all detected sources. Therefore, the 2BIGB catalogue brings an updated review for all the 925 3HSP-4FGL sources and describes the $\gamma$-ray spectrum of 235 3HSPs in addition to 4FGL (226 are entirely new and never reported in {\it Fermi}-LAT catalogues). 

This work is a followup from a series of previous efforts dedicated to the $\rm \gamma$-ray detection of blazars. The series includes the First version of the Brazil-ICRANet Gamma-ray Blazar catalogue  \citep[1BIGB, ][]{1BIGB}, the computation of the 1BIGB $\rm \gamma$-ray SEDs \citep{1BIGB-SED}, and the use of multifrequency seeds for the detection of LSP blazars \citep{LSPdetections}. To mention, the 1BIGB catalogue involved the analysis of 400 $\rm \gamma$-ray candidates from the 2WHSP catalogue \cite{2WHSP}, selecting only those with synchrotron peak flux $\rm \nu f _{\nu} \geq 10^{-12.1}$ ergs/cm$^2$/s and not yet detected by {\it Fermi}-LAT at the time. The 1BIGB study lead to the detection of 150 $\rm \gamma$-ray signatures with significance down to TS\,=\,10, from which 85 had TS\,>\,25.  

The 2BIGB catalogue goes beyond 4FGL for the description of EHSP \& HSP blazars since all 3HSP-4FGLs now have their $\rm \gamma$-ray spectrum modeled with 11 years of available data, in contrast to 8 years for 4FGL. Besides, this work provides a list of newly detected 3HSPs, which represents an improvement of $\sim$25\% for the number of EHSP \& HSP described in $\rm \gamma$-rays.   

%%%%%%%%%%%%%%%%%%%%%%%%%%%%%%%%%%%%%%%%%%%%%%%%%%%%%%%%%%%%%%%%%%%%%%%%%%%%%%%

\section{Motivations}

\subsection{An overview of the gal. \& extragal. gamma-ray sky}

A clear understanding of the $\rm \gamma$-ray sky depends on our ability to account for both galactic and extragalactic components and describe each of then in detail so that new features can be unveiled. 

%\subsection{The galactic gamma-ray content}

Concerning the galactic content, the {\it Fermi}-LAT mission was essential to map and model point-like sources such as pulsars, supernovae remnants (SNR), pulsar wind nebulae (PWN), globular clusters, etc ($\sim$239, 40, 18, 30, etc. objects according to the 4FGL catalogue), as well as a diffuse $\rm \gamma$-ray component produced by cosmic-ray interactions with the Milk Way gas \citep{Gal-DiffuseGamma-CR-Biswas2019}. Still, around 90 galactic sources remain unassociated but are likely related to SNR and PWN. 

In 2010, the finding of a bipolar component called ``Fermi Bubbles'' was unexpected \citep{Fermi-Bubble-SuMeng-2010}. It resembles a jet structure that emanates from the center of our galaxy, extending nearly 8 kpc perpendicular to the galactic plane, in what could be the result of past accretion process in the galactic center black hole \citep{Fermi-Bubble-origin-BHactivity-Ko2019}. This structure has well-defined edges in coincidence with X-ray signatures seen by the ROSAT mission in the 1990s \citep{GalCenter-Xray-Chimney-Ponti2019}, and a microwave excess in the same region known as ``WMAP haze'' \citep{WMAP-haze-Finkbeiner2004,WMAP-haze-Rubtsov2018}. Proper modeling of this component allowed to account for an excess diffuse emission, which covered a significant fraction of the $\rm \gamma$-ray sky. As a result, it improves the sensitivity for detecting new point sources in that region. 

Within the open question involving observations from {\it Fermi}-LAT, there is a long-standing debate related to the GeV excess at $\sim$1.5-3.0 kpc from the galactic center \citep[][]{GalCenter-Gamma-MilPulsar-Hooper2013,GalCenter-Gamma-DM-vs-Pulsars-Mirabal2013,DM-Gamma-GalCenter-Leane2019}. This feature could result from multiple unresolved point-like sources as millisecond pulsars \citep{Gamma-Diffuse-Pulsars-Calore2014}, or the integrated $\rm \gamma$-ray emission from our galactic dark matter (DM) halo via annihilation or decay channels \citep{DM-Gamma-GalCenter-Goodenough2009,DM-Gamma-GalCenter-Daylan2016,DM-Gamma-GalCenter-Karwin2017}. A combination of both contributions, including unresolved sources and DM signatures, are also considered in this contest.

%\subsection{The extragalactic gamma-ray content}

All current debate concerning the galactic $\rm \gamma$-ray content is crucial for proper modeling of the extragalactic content. The extragalactic emission ($\rm \gamma_{ExtraGal}$) is the remaining radiation when the total galactic contribution ($\rm  \gamma_{Gal}$) is subtracted from the entire $\rm \gamma$-ray content ($\rm \gamma_{Tot}$) seen by {\it Fermi}-LAT;  ($\rm \gamma_{ExtraGal}$\,=\,$\rm \gamma_{Tot} - \gamma_{Gal}$). The total $\rm \gamma_{ExtraGal}$ is usually referred to in the literature as extragalactic $\rm \gamma$-ray background \citep[EGB,][]{Ackermann2015-EGB}.

The EGB is a combination of well-resolved point-like sources with a diffuse -and isotropic- $\rm \gamma$-ray content. Detailed simulations of the CR interaction with the galactic disk shows that the isotropic diffuse component seen off the galactic plane must have extragalactic roots \citep{Gal-DiffuseGamma-CR-Biswas2019}. This component, usually called isotropic $\rm \gamma$-ray background \citep[IGRB,][]{Ackermann2015-EGB}, could well be the result of an unresolved population of faint $\rm \gamma$-ray emitters, including blazars \citep{EGBpaolo,1BIGB,DiffuseGammaBLlacs}. Nevertheless, the fact is, the community is still building an understanding of the total EGB composition.  

If one considers the IGRB as a sum of unresolved point-like sources plus an actual diffuse component, the increasing number of detectable blazars from 3LAC to 4LAC \citep{3lac,4lac} indicates an ever-shrinking space for the remaining diffuse component. Besides, direct searches for new $\gamma$-ray sources as in 1BIGB and 2BIGB (this work) proof the existence of a relevant underlying population of faint $\gamma$-ray emitters, with direct impact over the diffuse component intensity.

A diffuse component could be the signature of annihilation or decay processes associated with DM particles, happening throughout the universe. A clear understanding of IGRB content may allow us to probe DM parameters like the cross-section associated with processes (annihilation or decay), producing diffuse $\rm \gamma$-rays \citep{DM-Gamma-Constrains-Liu2017,DM-Gamma-Cohen2017,IGRB-DM-DiMauro2015,IsoDiffuseGammaBackGround}. Contributions to the EGB coming from misaligned AGNs \citep{Inoue-EGB-MAGN-2011,EGB-from-Radio-extrapolation-DiMauro2014} and Star Forming Galaxies \citep{EG_from-SFG-Storm2012} are also present. They certainly account for a significant fraction that could be well constrained via stacking analysis as in \cite{Stack-Gamma-Pop-Paliya2019}. Moreover, the IGRB could also contain a pure diffuse emission as a result of ultra-high-energy cosmic rays (UHECR) interaction with the extragalactic background light \citep{IGRB-from-URECR-Gavish2016}. In this context, the identification of faint $\gamma$-ray emitter is essential to resolve the diffuse content into point-sources and constrain the parameter space available for other possible contributions.  

%%%%%%%%%%%%%%%%%%%%%%%%%%%%%%%%%%%%%%%%%%%%%%%%%%%%%%%%%%%%%%%%%%%%%%%%%%%%%%%
%\section{Methodology}

%\textbf{Here we describe how to improve the description of the $\gamma$-ray sky by considering multifrequency information as a tool to unveil sources, which are considered faint in the energy window probed with {\it Fermi}-LAT but could turn-out to be more relevant at higher energies}. 

\subsection{Blazars as multifrequency seeds to investigate the gamma-ray sky}
\label{sec:blazars}

% 2BIGB Total: 1160 ; 925 in 4FGL + 235 new  
% 3HSP  Total: 2013 ; 1077 out of 4FGL
%Following the release of 4FGL catalogue, there still remains 1077 3HSP sources having no 4FGL counterpart.

\begin{figure}
   \centering
    \includegraphics[width=1.0\linewidth]{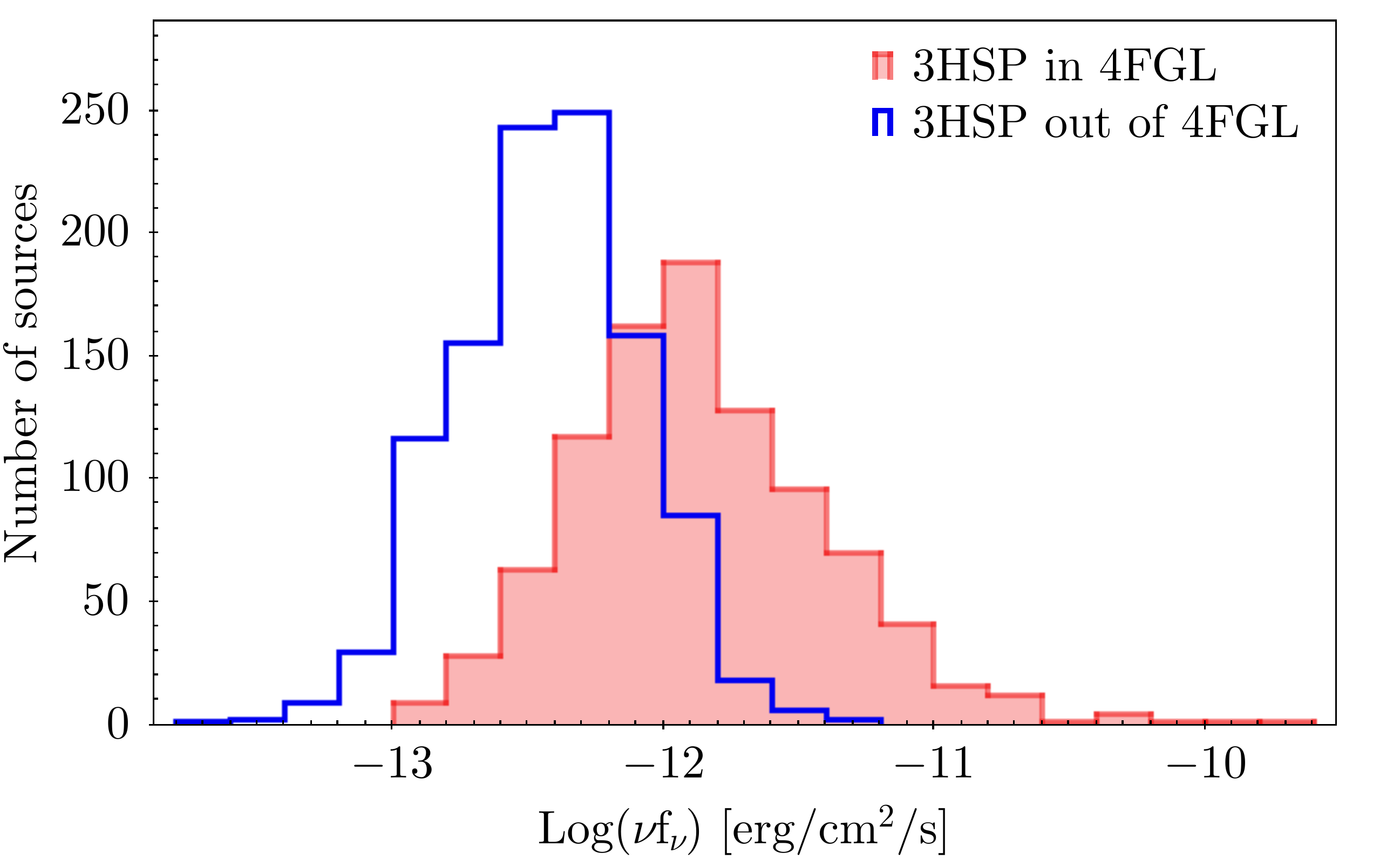}
     \caption{The distribution of log synchrotron peak flux log($\rm \nu f_{\nu}$) [erg/cm$^2$/s] for the 3HSP sources, with the 925 objects detected in 4FGL (full bars in light-red), and the 1088 undetected ones (blue line). The intersection between histograms suggests that multiple sources are close to the {\it Fermi}-LAT detection threshold.}
      \label{histofermi}
\end{figure}

The 3HSP catalogue lists a total of 2013 sources, and nearly half of them (1088) have no $\gamma$-ray counterpart according to the latest release of the {\it Fermi}-LAT point source catalogue, 4FGL. Fig. \ref{histofermi} shows the distribution of the log synchrotron peak flux log($\rm \nu f_{\nu}$) for the 3HSP detected and undetected $\gamma$-ray sources according to 4FGL. As seen, the 4FGL has unveiled most of the bright 3HSPs with log($\rm \nu f_{\nu}$)\,>\,-12.2 [ergs/cm$^2$/s], and many of the undetected 3HSPs have a synchrotron component as bright as the detected ones. 

Besides, the 1BIGB catalogue shows clearly how the `fraction of {\it Fermi}-LAT detected sources' varies according to the synchrotron peak flux log($\rm \nu f_{\nu}$), and that a dedicated search for new $\rm \gamma$-ray sources can improve the detected fraction down to the faintest blazars \citep[see Fig. 6 at][]{1BIGB}. Therefore, the overlap in Fig. \ref{histofermi} for detected and undetected sources at log($\rm \nu f_{\nu}$) [erg/cm$^2$/s] between -13.0 to -12.2, hints for a population of blazars that are just within reach of {\it Fermi}-LAT, close to the detectability threshold.

Those facts alone are compelling enough to motivate a dedicated {\it Fermi}-LAT analysis based on the entire 3HSP catalogue, using them as multifrequency seed positions to unveil new $\rm \gamma$-ray sources. Also, we rely on the availability of 11 years of {\it Fermi}-LAT data, in contrast to 8 years used to build the 4FGL catalogue. The 1BIGB catalogue was done over this same and straightforward idea, and lead to the detection of nearly 150 new $\rm \gamma$-ray sources, from which the 3FHL catalogue has confirmed 34 \citep{3FHL}, and a total of 99 are well within the 95\% error ellipses from 4FGL sources. 

%HERE IS A FIGURE FROM NEXT SECTION
\begin{figure*}
\includegraphics[width=0.90\linewidth]{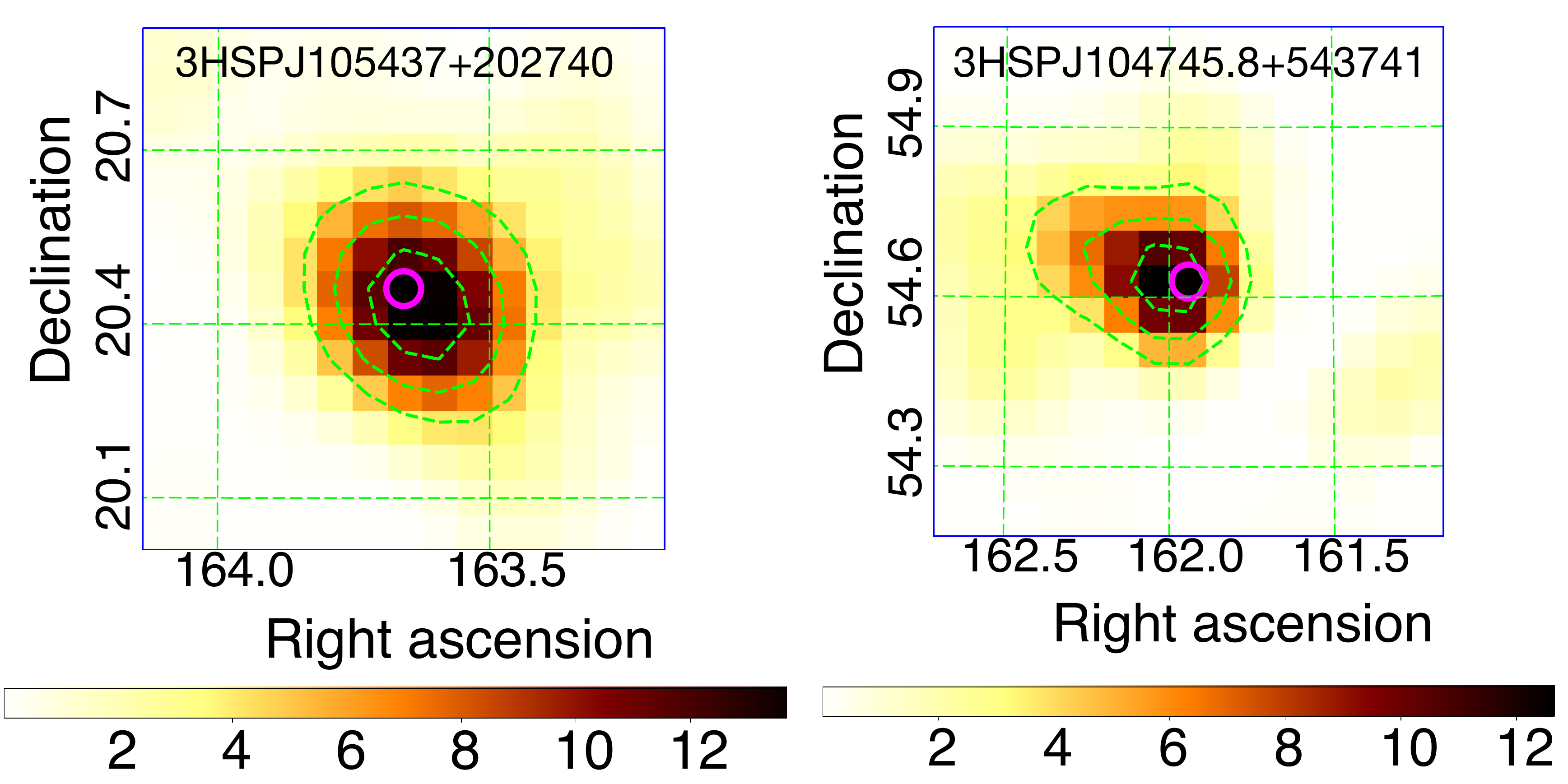}
\caption{The high-energy TS map center at 3HSP\,J105437.9+202740 (right) and 3HSPJ\,104745.8+543741 (left), integrating over 10.5 years of data only considering photons with E\,>\,2\,GeV. Those are examples of $\gamma$-ray signatures close to the detection threshold. The green dashed lines correspond to the 68\%, 95\%, and 99\% containment region for the $\gamma$-ray signature (from inner to outer lines). The TS values are smoother over two pixels with a 2$\sigma$ Gaussian function and are color-coded corresponding to the bottom legend.}
\label{fig:tsmap1}
\end{figure*}

The method applied to build the 2BIGB catalogue is a follow-up from previous works \citep{1BIGB-SED,1BIGB,LSPdetections} where multifrequency data is used as a complementary tool to select seed positions in the sky for further analysis with the Fermi Science Tools. Mostly, we reproduce what the Cerenkov Telescope community already does in search of new TeV sources, which means, to actively aim to seed-candidates selected from multifrequency data. One of the latest detections from the Major Atmospheric Gamma Imaging Cherenkov - MAGIC \citep{2WHSP-TeV-Detected-2019} is a clear example of that, where a new TeV signature is confirmed when aiming at 2WHSP\,J073326.7+515354, which is well characterized at lower energies and previously selected as a promising TeV target from multifrequency considerations.  

The search for new $\gamma$-ray sources as in 1 \& 2BIGB catalogues can also contribute to studies of astrophysical neutrinos in connection to $\rm \gamma$-ray blazars. Recently in \cite{Second-Neutrino-Blazar-GB61040-2019,IceCube-NeutrinoTrack-2018}, the authors dedicate attention to the identification and modeling of dim $\rm \gamma$-ray sources in regions associated with IceCube muon-track events. With no doubt, proper identification of faint GeV-TeV sources is of high relevance. Also, a source can be weak in terms of photon counts but still relevant in the VHE window, especially for cases where the {\it Fermi}-LAT only probes the rising tail of the higher-energy SED peak.

\section{Methodology}

\subsection{The Gamma-ray data analysis}
\label{section:second}

For the broadband data analysis, we use the Fermi Science Tools version v11r5p3 and Pass 8 data \citep[P8R3,][]{P8R3-2018,PASS8} corresponding to 11 years of observations, from Aug 2008 to Aug. 2019. We build an all-sky pre-selection of photons with energy between 300\,MeV up to 900\,GeV, and use the clean-event type (evtype=256) with its corresponding instrument response function (IRF) P8R3-CLEAN-V2. The use of clean-event type is desirable in our case since it provides an improved lower-level background at E$>$3\,GeV; therefore, more sensitive to hard spectrum sources as the HSP blazars. The all-sky photon file (our pre-selection) speeds up the gtselec routine, which is the first step of the analysis, and repeated for each object. 
%The current software is version v11r5p3, released Feb 15, 2018
%The Clean event type, the residual background changes with the event class, 

After the pre-selection, we set a broadband binned analysis in the 500\,MeV to 500\,GeV energy band, considering a region of interest (ROI) of 15$^\circ$ around the position of a 3HSP $\gamma$-ray candidate. The investigation for each 3HSP object is independent, meaning that each analysis probe only one $\gamma$-ray candidate at a time. We accomplish the coverage over the entire 3HSP catalogue by parallel computing in computer clusters as Planck and Feynman from CCJDR Unicamp BR, and Joshua from ICRANet IT. 

The cut at 500\,MeV serves two purposes: On one hand, it is meant to improve the computational time of the analysis, which gets heavier when considering low-energy photons (those are relatively more abundant). On the other hand, it improves the robustness of the results while avoiding low-energy photons that have the largest point spread function (PSF) in the database. The low-energy cut is desirable in our case given the focus on EHSP \& HSP sources, which are characterized by relatively hard $\rm \gamma$-ray spectrum and more relevant at the highest energies. Also, cutting off the low-energy photons let the analysis free of computing energy dispersion at energy levels lower than 300\,MeV, which would significantly increase the computational burden. 

The cut at 500\,GeV follows recommendations from the {\it Fermi}-LAT team for broadband analysis (in ``data preparation" section at \url{https://fermi.gsfc.nasa.gov/}). Also, the 4FGL paper \citep[their sec. 3.2,][]{4FGL} comment on the fact that for hard sources, integrating up to energies of 1\,TeV in a broadband analysis systematically increases the uncertainty for parameters such as the photon and energy fluxes. Therefore, with a cut at 500\,GeV the analysis avoids uncertainties for the broadband flux, and prevent spurious detections that could arise from poor modeling of the VHE diffuse background. 

%[Removed this. Seems to be old comment if Fermi-LAT pages. Now, the Isotropic file is adjusted till 1TeV and is extrapolated up to 3TeV following the power-law trend. ]
%This takes into account that the galactic diffuse emission is only modeled for the 58.5\,MeV to 513\,GeV range, which mostly impact sources close (or within) the galactic plane. We should care about that since the 3HSP catalogue probe sources down to the galactic latitude |b|>10$^{\circ}$, and some extra 3HSPs are within the galactic plane; those were included in 3HSP for having clear blazar-like SED in connection to 3FHL and 4FGL sources \citep{3HSP}.

A zenith angle-cut of 105$^\circ$ is used to avoid contamination with Earth's limb $\rm \gamma$-ray photons, which are induced by cosmic-ray interactions with the atmosphere and are known as a strong source of background for the low-energy band of {\it Fermi}-LAT. With the gtmktime routine, we generate a list of good time intervals, selecting events with flags (DATA-QUAL$>$0) and (LAT-CONFIG==1), which guarantees to only consider data acquired in normal science data-taking mode. Then, with the gtbin routine, we generate counts maps (CMAP) and counts cubes (CCUBE) of $300 \times 300$ and $210 \times 210$ pixels with $0.1^\circ$/pixel, respectively. The CCUBE is a series of CMAPs, each one having photons within 37 logarithmic equally-spaced energy bins along 0.5-500 GeV. The livetime cube is build with the gtltcube routine, selecting cos(theta)=0.025$^\circ$ as recommended.% in {\it Fermi}-LAT analysis threads). 

%The fourth column shows the redshift reported in 3HSP catalogue, with flag (1) for firm value and flag (5) for photometric estimate by fitting a Giant Elliptical galaxy template.
%LONG-TABLE-1
\begin{table*}
\caption{Power-law model for the top 10 2BIGB-3HSP new $\rm \gamma$-ray sources. Note that a complete table with all the 1160 2BIGB source is available in the on-line version of this paper, and also at public repositories; Github \url{https://github.com/BrunoArsioli}. The first three columns show the 2BIGB names, right ascension R.A. and declination Dec. in degrees (J2000), respectively. The fourth column shows the reported redshifts from literature  \citep{bllacz2,pita2013,pks1424p240HighZ,PG1553,shaw2,masetti2013,bll,5BZcat-V5}, with a right uppercase flag where (1) correspond to cases with a robust redshift value, (2) for values reported as uncertain, (3) for lower limits reported in 3HSP catalogue, and for a photometric estimate by fitting a Giant Elliptical host galaxy template we have (4) when a featureless optical spectrum is available, and (5) when no optical spectrum is available but only photometric data points. The parameters reported in columns are the normalization $\rm N_0$ (eq. \ref{eq:powerlaw}), which is given in units of ph/cm$^{2}$/s/MeV; the photon spectral index $\rm \Gamma$; the column Flux shows the photon counts integrated over 0.5-500\,GeV in units of ph/cm$^2$/s; the pivot energy E$_0$; and the TS is the Test Statistic value.}
\label{table:2BIGB-10yrs-slim}
{\def\arraystretch{1.3}
 \begin{tabular}{llrcccccc}
\hline
2BIGB name  & R.A. (deg)  &  Dec. (deg)  &  z  &  $\rm \Gamma$ & N$_0$ (10$^{-15}$)  & Flux$_{0.5-500\,GeV}^{( \times 10^{-11})}$ & E$_{0}$ (GeV) & TS \\
 \hline
  210415.9+211808 &  316.06633 &  21.30228   &  $^{(5)}$0.36   & 1.91$\pm$0.14 &    13.2$\pm$2.3   &  59.6$\pm$17.5  & 5.0   & 122.3  \\       
  235955.0+314600 &  359.98042 &  31.76667   &  $^{(5)}$0.33   &  1.85$\pm$0.12 &   8.54 $\pm$1.38  & 35.6$\pm$ 7.0   &  5.0  &  96.8  \\      
  111717.5+000633 &  169.32308 &  0.10931    &  $^{(1)}$0.451  &  1.76$\pm$0.13 &   7.85 $\pm$1.50  & 29.6$\pm$ 7.4   &  5.0  &  74.3  \\      
  230848.7+542611 &  347.20308 &  54.43644   &  -              &  1.71$\pm$0.12 &   2.55 $\pm$0.47  & 30.1$\pm$ 7.8   & 10.0  &  66.9  \\     
  083015.1-094455 &  127.56308 &  -9.74883   &  $^{(5)}$0.5    &  1.85$\pm$0.15 &  15.8 $\pm$3.7    & 25.6$\pm$ 7.4   &  3.0  &  53.1  \\      
  225613.3-330338 &  344.05546 &  -33.06064  &  $^{(1)}$0.243  &  1.99$\pm$0.17 &  15.9 $\pm$4.4    & 28.5$\pm$10.2   &  3.0  &  51.6  \\
  030103.7+344101 &  45.26558  &  34.68367   &  $^{(1)}$0.246  &  2.13$\pm$0.17 &  21.5 $\pm$4.1    & 44.4$\pm$11.6   &  3.0  &  49.8  \\   
  235917.0+021520 &  359.8210  &  2.25564    &  -              &  1.85$\pm$0.14 &   5.40 $\pm$1.20    & 22.5$\pm$ 6.1   &  5.0  &  48.3  \\      
  173044.8+380454 &  262.68663 &  38.08192   &  $^{(5)}$0.22   &  1.79$\pm$0.17 &   4.64 $\pm$1.18  & 18.0$\pm$ 6.6   &  5.0  &  40.9  \\      
  101724.4+253956 &  154.35158 &  25.66556   &  $^{(1)}$0.417  &  1.40$\pm$0.19 &   0.96 $\pm$0.30  & 7.54$\pm$ 3.14   & 10.0  & 40.6  \\     
 \hline
 \end{tabular}
 }
\end{table*}

The $\gamma$-ray sky models are build with the make4FGLxml.py\footnote{The makeFGLxml.py is a python routine written by T. Johnson, 2015, and provided by the {\it Fermi}-LAT team as a user contribution tool.} script, which includes information on all point and extended sources in the ROI region, together with the galactic and extragalactic diffuse components. For the point sources, we use the latest version of the {\it Fermi}-LAT catalogue gll-psc-v20.fit, which we call 4FGLv20. For the diffuse galactic background model and the isotropic component, versions gll-iem-v07.fit and  iso-P8R3-CLEAN-V2-v1.txt are used, respectively. All sources within 8.0$^\circ$ from the center ROI have spectral parameters -normalization and photon index- set free to vary, which is essential to adapt the 4FGL models (based in eight years) to the 11 years of integration time used in this work. This is necessary for a proper description of the ROI region and to avoid spurious signatures associated with our candidates. 

Then, a model map is prepared with the gtsrcmaps routine, which holds information about all spectral components and sources within the ROI region. Each 3HSP object is placed in the ROI model as a point-like source, and described with a power-law type of spectrum as follows: 
\begin{equation} \rm
\hspace{85pt}\frac{dN}{dE} \, \text{=} \, N_0 \; \left( \frac{E}{E_0} \right)^{-\Gamma} \;,
\label{eq:powerlaw}
\end{equation} 
where $\rm N_0 $  is the normalization constant (prefactor) in units of photons/cm$\rm ^2$/s/MeV and represents the flux density calculated at the pivot-energy $\rm E_{0}$, with $\rm \Gamma$ to represent the photon spectral index.  

In cases where a 4FGL is already counterpart of the 3HSP, we remove the 4FGL form the ROI model and place a new $\gamma$-ray candidate using the 3HSP position as seed. For all cases, the position associated with the $\gamma$-ray source (or candidate) is fixed, therefore lowering the uncertainty related to source position. This is the main advantage of considering multifrequency data to select seed positions. Given that the 2BIGB analysis avoids two degrees of freedom concerning a blind search, we can lower the detection threshold. %down to TS 9 \citep{mattox}   

In fact, the detection of a new class of $\rm \gamma$-ray emitter demands large significance, given the analysis cannot avoid the uncertainty on the source position when deduced from $\gamma$-ray information only. From \cite{Mattox1996}, the $\chi^2$ distribution with four degrees of freedom shows that a Test Statistic (TS) of 25 corresponds to a detection with $\sim 4\sigma$, just as the detection threshold defined in the 4FGL catalogue\footnote{The Test Statistic parameter is defined as $ \rm -2 ln \left( L_{\left( no-source \right)} \div L_{\left( source \right)} \right) $, where $\rm L_{ \left( no-source \right)}$ is the likelihood of observing a given photon count only due to background (null-hypothesis), and $\rm L_{\left( source \right)}$ is the likelihood of observing a given photon count considering a source exists in a particular position \citep{Mattox1996}.}. In this case, there are two degrees of freedom related to source position (R.A. and Dec.), and another two related to the normalization N$_0$ and the photon spectral index $\Gamma$, which are free parameters of the power-law model.  

%Following \cite{Mattox1996} (their sec. 3.2), a $\rm \gamma$-ray signature with the statistical significance of the order of $\rm \sim 3\sigma$ is sufficient for the detection of blazars, because blazars are the most relevant class of extragalactic gamma-ray emitter. A lower detection threshold is acceptable, given the analysis has less two degrees of freedom (R.A. and Dec. are set as fixed to avoid the uncertainty related to the source location). 

%[We COULD SAY] 
Following \cite{Mattox1996} (their sec. 3.2), a $\rm \gamma$-ray signature with statistical significance of the order of $\rm \sim 3\sigma$ is sufficient for the detection of blazars, because blazars are the most relevant class of extragalactic gamma-ray emitter. A lower detection threshold is acceptable, given the analysis has less two degrees of freedom (R.A. and Dec. are set as fixed to avoid the uncertainty related to the source location). For a broadband analysis with two degrees of freedom, a $\rm 3\sigma$ detection corresponds to TS$\sim$13. In our analysis, we lower the threshold down to TS = 9 ($\rm \sim2.3 \sigma$) to include faint detections for future follow-up\footnote{The correspondence between the TS value, the number of degrees of freedom, and the significance of a $\gamma$-ray signature can be calculated according to \url{https://github.com/BrunoArsioli/TS-DegFreedom-Sigma-relation-FermiLAT}.}. Given that we investigate $\sim$2000 seed positions, the total number of spurious detections expected is of the order of $ \rm  \frac{2000}{10^{2.3}} \sim 10 $. Moreover, the number of spurious detection might be even lower, given that we evaluate the TS distribution of all new detections, and actively removed all likely-spurious cases which did not reveal itself as a point-like source.  

%Following \cite{Mattox1996} (their sec. 3.2), a $\rm \gamma$-ray signature with the statistical significance of the order of $\rm 3\sigma$ is sufficient for the detection of blazars, because blazars are the most relevant class of extragalactic gamma-ray emitter. A lower detection threshold is acceptable, given the analysis has less two degrees of freedom. That is what we adopt in this work since R.A. and Dec. are set as fixed, avoiding the uncertainty related to the source location. 

Since the prefactor term $\rm N_{0}$ corresponds to the flux density at the pivot energy $\rm E_{0}$, we scan over the $\rm E_{0}$ space, running the likelihood analysis with $\rm E_{0}$ set as fixed for 1\,GeV, 3\,GeV, 5\,GeV and 10\,GeV to evaluate the best condition for modeling each source. This particular step allows to select a pivot-energy that minimizes the error associated to $\rm N_{0}$. For cases where the 3HSP has a 4FGL counterpart, the pivot-energy corresponds to  4FGLv20. We perform the analysis with $\rm E_{0}$ set as fixed so that $\rm N_{0}$ and $\rm \Gamma$ could be estimate with much lower uncertainty. 

The likelihood analysis goes through two steps with the gtlike routine, following recommendations from the {\it Fermi}-LAT analysis threads. First, the gtlike runs with the fitting optimizer DRMNFB to generate enhanced sky models, better describing all sources with free parameters. After this first interaction, the model map is rebuild with the enhanced source-input list, and feed the gtlike routine, now with the NEWMINUIT optimizer, to generate the final model.

\subsection{Building the 2BIGB catalogue}

With the results of our likelihood analysis, we set the requirement of TS>9 to select the preliminary list of sources with significant $\gamma$-ray signature. This list included all 925 3HSPs with a counterpart in 4FGL, which are considered as firm detections. It also included $\sim$270 cases with no counterpart in 4FGL, and that could turn out to be new $\gamma$-ray detections. Therefore, in addition to the TS requirement, we build high-energy TS maps with an energy cut of E$>$2\,GeV for all cases with no counterpart in 4FGL. 

A TS map is a grid of pixels where the existence of a point-like $\gamma$-ray source is tested separately in each point of the grid. The grids have $\sim$18$\times$18 pixels with 0.06$^\circ$/pixel. If a pixel match with the position of a point source, the TS signature is maximized (large TS values), and if there is a slight offset between the pixel and the exact position of the point-source, the reported TS values are depleted. Therefore, a point source is expected to reveal itself as a smooth Gaussian-like distribution of TS values around a TS peak, as seen in Fig. \ref{fig:tsmap1}  \citep[build with DS9 software,][]{ds9}. With this procedure we confirmed 235 sources out of the 270 candidates, clearly showing point-like signatures emerging from a smooth and low TS background as in Fig. \ref{fig:tsmap1}.

Finally, the 2BIGB catalogue has a total of 1160 sources with a broadband excess signature corresponding to TS\,>\,9, from which 925 have a 4FGL counterpart, and 235 are new-detections concerning 4FGL. Table \ref{table:2BIGB-10yrs-slim} shows a sub-sample of 10 new-detection with the most significant TS values. A list with the total 1160 sources is available in the on-line version of this paper, and also in public repositories; Github \url{https://github.com/BrunoArsioli}. For simplicity, we define the acronyms: 

\begin{itemize}
    \item 3HSP-2BIGB corresponds to the total 1160 3HSP sources detected in our broadband analysis over 11 years of {\it Fermi}-LAT observations, along the 0.5-500\,GeV energy band.
    \item 3HSP-4FGL corresponds to the 925 3HSPs sources that already have a 4FGL counterpart, as of version gll-psc-v20.
    \item 2BIGB$_{new}$ refers to the 235 2BIGB sources which are new with respect to 4FGL. Moreover, 226 were never reported in previous $\gamma $-ray catalogues, including 1-2-3-4FGL \& 1-2-3FHL \citep{1FGL,2FGL,3FGL,4FGL,1FHL,2FHL,3FHL}.
\end{itemize}

\subsection{The gamma-ray Spectral energy distribution}
\label{section:fourth}

The 2BIGB catalogue presents a SED description for all its 1160 sources, with estimated flux values (or Upper Limits) for ten energy bands, with data-points ranging from 1\,GeV to 170\,GeV. Note that the broadband analysis is updated to the latest version of the point-source catalogue 4FGLv20 (gll-psc-v20.fit) and integrated over 11 years. However, the $\gamma$-ray SED is built considering 10.5 years of observations, and having 4FGLv19 (gll-psc-v19.fit) as the library of point-sources to build the sky models.

In {\it Fermi}-LAT catalogues, the energy bins are defined by dividing the broadband window in equally spaced logarithmic values, which tends to produce mostly UL values in faint $\gamma$-ray sources. However, in the 2BIGB catalogue, the fluxes are estimated by integrating over superposed energy windows (as listed in Table \ref{table:E-bin}), which are larger than equally spaced logarithmic bins. By doing so, we incorporate valuable broadband information into the SED description, following the same approach as in previous work \citep{1BIGB-SED}, and allowing us to extract SED information even for the faint sources. To fit each SED data-points, both N$\rm _0$ and $\rm \Gamma$ are left free to vary, setting an independent power-law model for each energy bin. This way, the SED final shape is more sensitive to the actual spectral curvature, and not bound to the broadband power-law modeling (500\,MeV to 500\,GeV). In the case of 4FGL, for example, only $\rm N_0$ is allowed to vary (fit) in each energy bin, and the $\rm \Gamma$ parameter is set fixed according to its broadband power-law modeling \citep{4FGL}.    

\begin{table}
	\centering
	\caption{The column `Integrated over' shows the definition of the energy bins used for the spectral analysis. We estimate the flux density $\rm N_0$ [ph/cm$^2$/s/MeV] for each pivot-energy $\rm E_0$ within its corresponding bin, and the analysis-type binned/unbinned is also listed.}
	\label{table:E-bin}
	{\def\arraystretch{1.1}
	\begin{tabular}{l|l|l} % four columns, alignment for each
		\hline
		E$_0$ [GeV] & Integrated over [GeV]  &  Analysis-Type \\
		\hline
 		1.0   & 0.7 - 5.0        & binned   \\
		1.7   & 0.8 - 10.0       & binned   \\
		3.0   & 0.9 - 17.0       & binned   \\
        5.0   & 1.0 - 30.0       & binned   \\
		10.0  & 1.7 - 50.0       & binned   \\
		17.0  & 3.0 - 100.0      & unbinned \\
        30.0  & 5.0 - 170.0      & unbinned \\
		50.0  & 10.0  - 300.0    & unbinned \\
		100.0 & 30.0  - 500.0    & unbinned \\
		170.0 & 50.0  - 800.0    & unbinned \\ 
		\hline
	\end{tabular}}
\end{table}

To build a spectral description for all 1160 2BIGB sources represents a massive load of nearly 12k likelihood analysis to complete, divided between binned and unbinned for the lower and higher energy bands, respectively. Table \ref{table:E-bin} lists the pivot-energies ($\rm E_0$) at which the fluxes are calculated, by integrating over the correspondent energy band. We choose those E$_0$ values to be close to logarithmic equally spaced SED data points when plotting on commonly used spectrum planes for blazars, like log($\nu$) vs. log($\nu f _{\nu}$). Therefore E$_0$ increases from 1\,GeV up to 170\,GeV with increments of the order of $\sim$10$^{0.25}$.

For the highest energy bands (17\,GeV up to 170\,GeV), we applied unbinned analysis, which is more compute-intensive but better suited for cases where the photon count is low. This approach follows from the 1BIGB paper data analysis, which showed that a binned analysis could lead to an underestimated flux at 100\,GeV when compared to sources detected in 2FHL and 3FHL. Although a cut at 500\,GeV was set for the broadband analysis, the use of photons with E$>$500\,GeV is valuable to estimate the flux at the very-high-energy bins. Especially for bright and extreme sources, which can be detected by {\it Fermi}-LAT up to the energy level of 100\,GeV and larger, we do integrate from 50\,GeV up to 800\,GeV to build the largest energy data-point at 170\,GeV. In this case, the uncertainties related to VHE photon/energy fluxes are contained in the last bin and adequately accounted for by the flux error estimate.   

For the data analysis, all sources within 8$^\circ$ from the 2BIGB position have free spectral parameters to fit, just as in the broadband analysis. Given the choice of $\rm E_0$ for each bin, we get a reliable calculation for the normalization parameter $\rm N_0$, which is the differential flux at that specific energy. In other words, the SED data points are direct outputs from the likelihood analysis, only by converting $\rm N_0$ [ph/MeV/cm$^2$/s] to flux [erg/cm$^2$/s];  Flux$\rm _{\left(E_0 \right)}$\,=\,$ \rm N_0 \cdot E_0^2 \cdot 1.602 \times 10^{-6}$, where the factor 1.602$\times$10$^{-6}$ is to convert MeV to erg. Whenever a band reaches TS\,<\,6.0, an upper limit value is assigned. Especially, for the largest energies of 100\,GeV and 170\,GeV, we raise the UL threshold to TS\,<\,10.0 to be more conservative and only report on confident VHE flux estimates. For the upper limit values, we use the Broadband-Flux-Sensitivity-l0-b30 (for P8R3-V2 and 10 years of exposure), which are conservative compared to larger galactic latitudes\footnote{The {\it Fermi}-LAT performance is described at  \url{https://www.slac.stanford.edu/exp/glast/groups/canda/lat_Performance.htm}.}.

The SED data points from the 2BIGB catalogue are suitable to identify the mean spectral shape over time, given the large integration window of 10.5 years, but also smooth in terms of energy, given the superposed energy bins. Fine spectral structure derived from this data-set in combination with TeV data from Cherenkov Telescopes should be considered with care and case by case, especially to evaluate the influence of spectral variability along time. If not properly accounted, short-time flaring events can mimic spectral structure when integrating over a large time window.

\section{Results \& Discussion} 

\subsection{A comparison between 2BIGB and 4FGL}

% THIS IS A FIGURE FROM NEXT SECTION 
\begin{figure}
\includegraphics[width=1.0\linewidth]{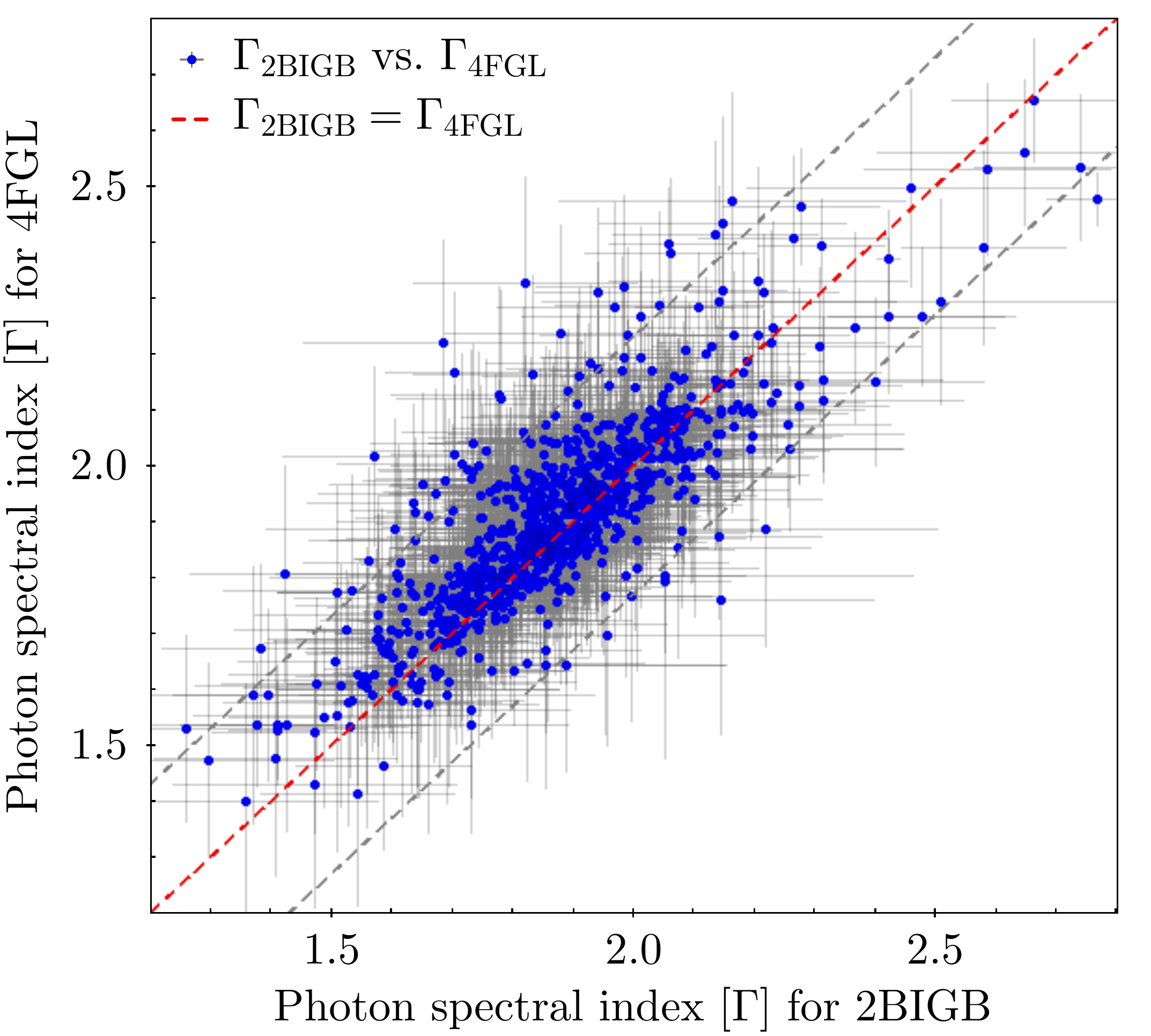}
\caption{Scatter plot comparing the photon spectral index for the 925 2BIGB sources with counterparts in 4FGL (the 3HSP-4FGL sources). Dashed red represent the $ \rm \Gamma_{2BIGB}$=$\rm \Gamma_{4FGL}$ line, and dashed gray represents the 2$\sigma$ confinement region.}
\label{fig:slopes}
\end{figure}

% THIS IS A FIGURE FROM NEXT SECTION 
\begin{figure}
\includegraphics[width=1.0\linewidth]{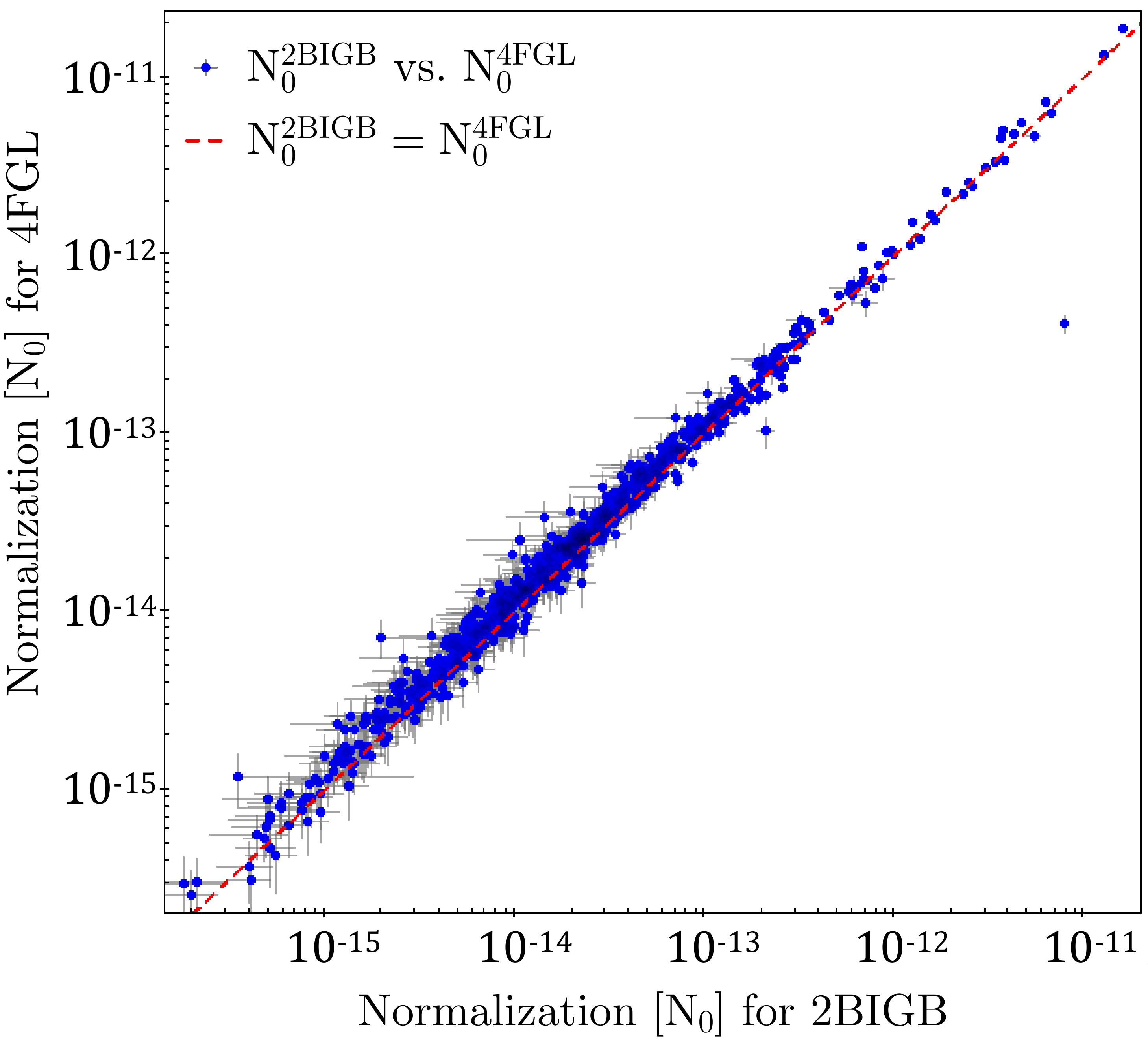}
\caption{Scatter plot comparing the normalization parameter $\rm N_0$ for the 925 3HSP-4FGL sources with counterparts in 2BIGB. Red dashed represent the $ \rm N_0^{2BIGB}$=$\rm N_0^{4FGL}$ line, which shows good agreement along five decades.}
\label{fig:n0}
\end{figure}

We chose to investigate the entire 3HSP catalogue with {\it Fermi}-LAT to update the power-law fitting parameters for all the 925 3HSP-4FGL sources, now based on 11 years of observations. Besides, it is possible to compare all 3HSP-4FGL with the 3HSP-2BIGB sources, and evaluate the agreement between the main fitting parameters $\rm \Gamma$ and $\rm N_0$. Fig. \ref{fig:slopes} shows the scatter plot of the photon spectral index from 3HSP-4FGL ($\rm \Gamma_{4FGL}$) versus the 3HSP-2BIGB sources ($\rm \Gamma_{2BIGB}$). And Fig. \ref{fig:n0} shows the normalization parameter $\rm N_0^{4FGL}$ vs. $\rm N_0^{2BIGB}$. In both cases, we consider all 925 3HSP-4FGL sources that have a 2BIGB counterpart, and compare their power-law fitting parameters. A certain degree of scattering is expected, as seen in Fig. \ref{fig:slopes} and \ref{fig:n0}.

Note that the 2BIGB broadband analysis covers the 500\,MeV to 500\,GeV energy window, whereas 4FGL covers 50\,MeV up to 1\,TeV. The 2BIGB avoids the lower energy band (from 50\,MeV to 500\,GeV), which is a source of uncertainty given the broader PSF of low energy photons. Also, the 2BIGB incorporates an additional three years of observations concerning 4FGL, and intrinsic variability becomes another source of data spread. The scattering for the $\rm \Gamma$ plane is well confined within 2$\sigma$ deviation, and there is good agreement between 2BIGB and 4FGL for the $\rm N_0$ parameter along five decades in flux\footnote{The single outlier is 3HSP\,J224910.7-130002, which is detect in 2BIGB with TS$\sim$15000, while detected with TS$\sim$121 in 4FGL, therefore a source which probably went flaring between Oct 2016 to Oct 2019.}. We conclude that our 2BIGB analysis is robust and consistent with the results reported in 4FGLv20, and this agreement supports the validity of our 235 new detections (2BIGB$_{new}$) associated with 3HSP sources. 

\subsection{Detectability and sensitivity limit according to FOM}

\begin{figure}
\includegraphics[width=1.0\linewidth]{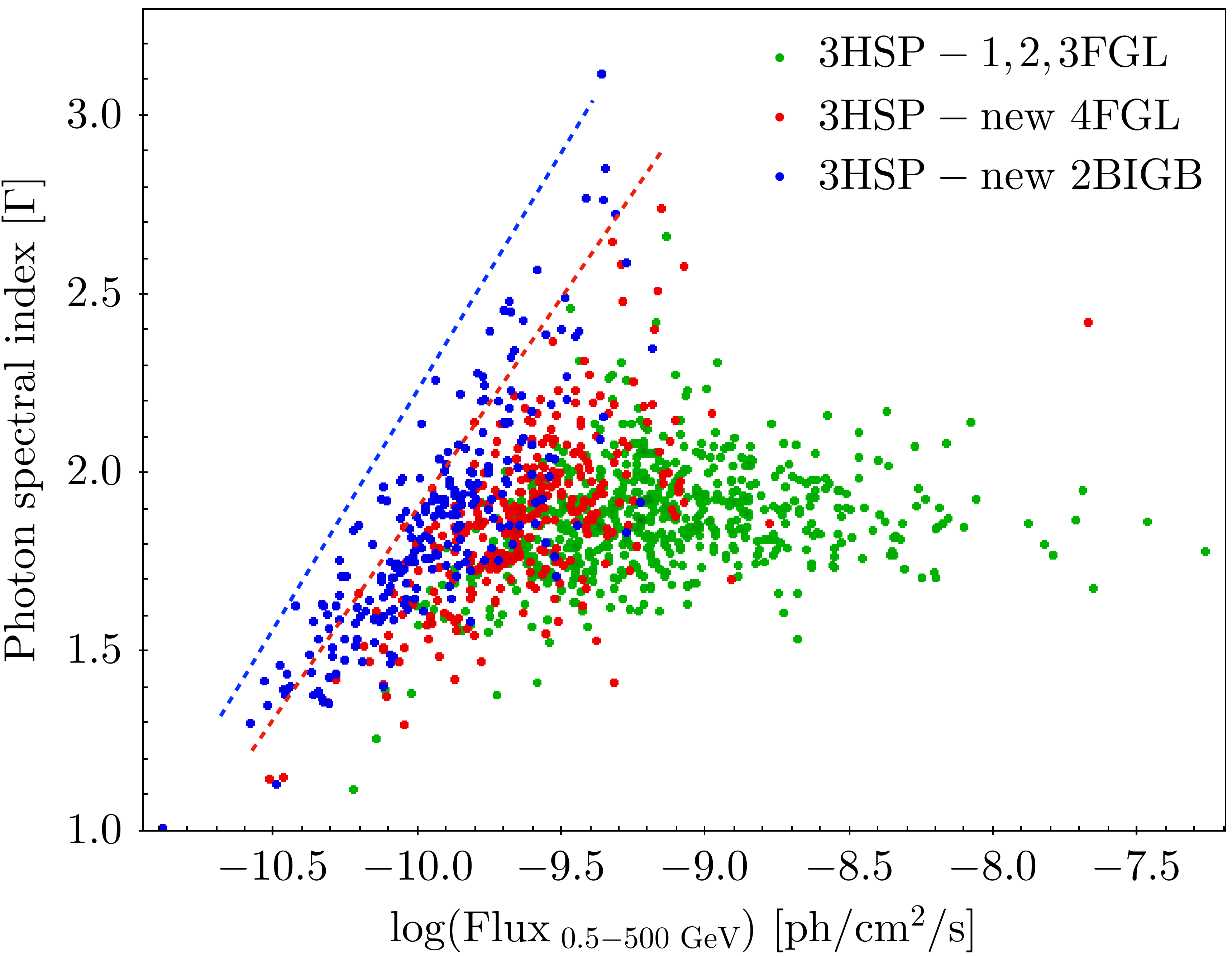}
\caption{The photon spectral index ($\rm \Gamma$) versus log(flux) for the energy range 0.5-500\,GeV. Green, red, and blue colors represent respectively 3HSP sources with counterparts from 1,2,3FGL, only 4FGL, and the new detections from the 2BIGB catalogue. The dashed lines represent a qualitative flux limits} reached by the 4FGL (red) and the 2BIGB (blue) catalogues.
\label{fig:fluxlimit}
\end{figure}

The Fig. \ref{fig:fluxlimit} shows the  photon spectral index ($\rm \Gamma$) versus the integral flux (0.5-500\,GeV) from where it is possible to perceive the improvement in sensitivity limit when going from the 3FGL setup to 4FGL, and finally to 2BIGB. Besides, the overlap between 2BIGB and 4FGL sources shows the improvement in completeness close the faint end of the 3HSP-4FGL sample. The dashed lines are qualitative representations of the flux-limit and its dependence with respect to $\Gamma$, for both the 3HSP-2BIGB and 3HSP-4FGL samples. The improvements seen for the 2BIGB catalogue result from the extra three years of data, the use of multifrequency information, and the adoption of a lower detection threshold to search for new sources.  

%Now, we read the Figure of Merit (FOM) parameter from the 3HSP catalogue, and calculate the fraction of sources in each FOM bin that are detected in 4FGL and in 2BIGB. 

Each 3HSP source has an associated Figure of Merit (FOM) parameter. The Figure of Merit is defined as a ratio between the synchrotron peak flux $\rm \nu f_{\nu} $ of a given source, and the faintest $\rm \nu f_{\nu} $ associated to a TeV detected HSP source ($\rm \nu f_{\nu} $\,=\,2.5$\times$10$^{-12}$ erg/cm$^2$/s, log($\rm \nu f_{\nu} $)=-11.6). Therefore, the FOM is a parameter connected to the likelihood of TeV detectability \citep[see][for more details]{3HSP} and here we show it indeed works as a reliable proxi for the $\gamma$-ray detectability with {\it Fermi}-LAT. 

\begin{figure}
\includegraphics[width=1.0\linewidth]{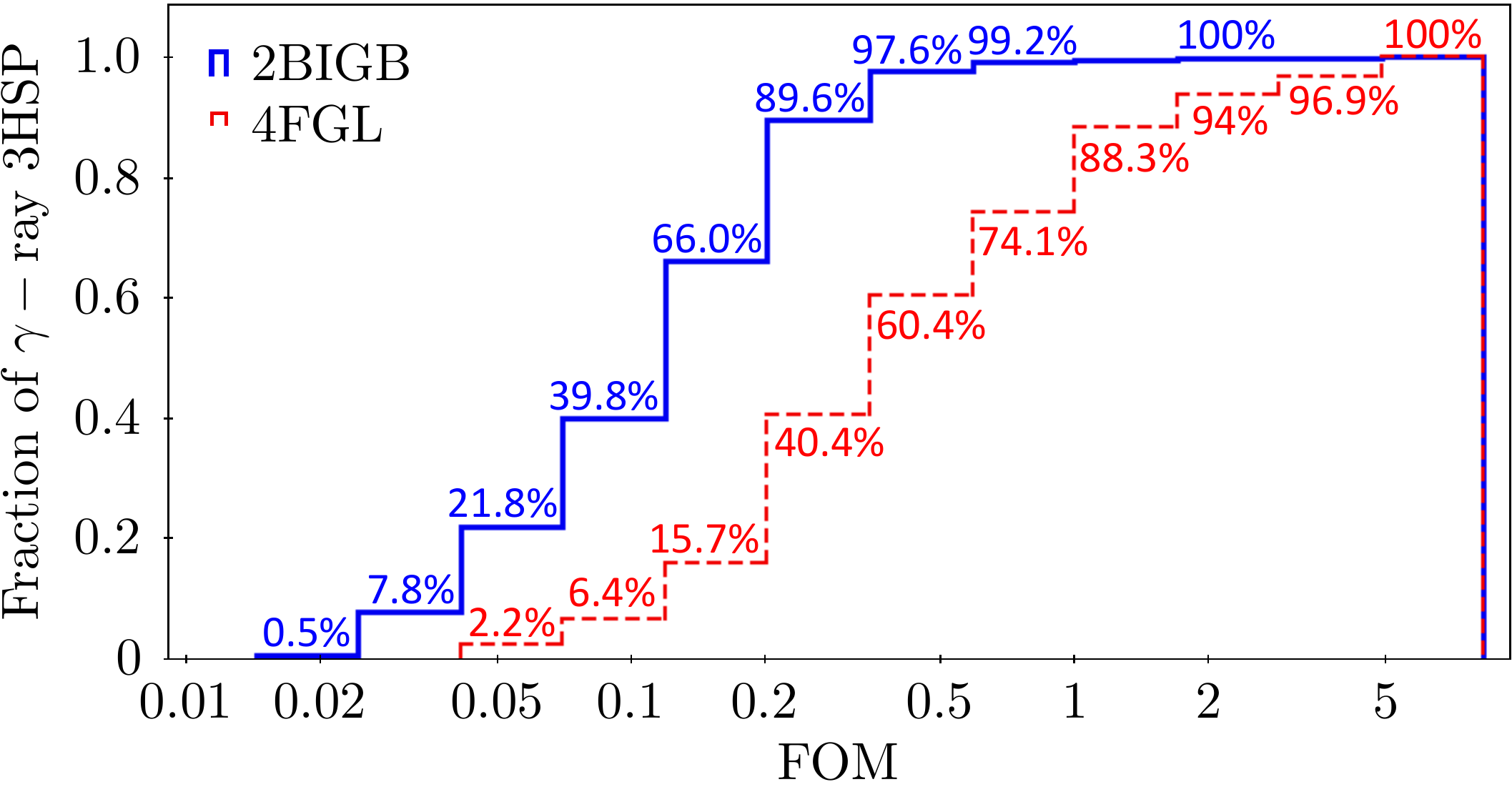}
\caption{Histogram for the cumulative fraction of 3HSP sources detected in $\rm \gamma$-rays for each FOM logarithmic bin. The 2BIGB sources are shown in blue and the 4FGL in red. Note that the last bin to the right condense all 3HSP sources with FOM$>$5. Here, FOM\,=\,log($\nu f_{\nu}$)/-11.6 [erg/cm$^2$/s].}
\label{fig:FOM}
\end{figure}

Figure \ref{fig:FOM} shows that the fraction of $\gamma$-ray detected sources increases with the FOM value, for both 2BIGB and 4FGL catalogues. Moreover, Fig. \ref{fig:FOM} shows that the 2BIGB catalogue brings an improvement in sensitivity concerning 4FGL, contributing to the $\gamma$-ray detection of a larger fraction of blazars in every FOM bin. 

\subsection{The photon spectral index \& flux distribution}

\begin{figure}
\includegraphics[width=1.0\linewidth]{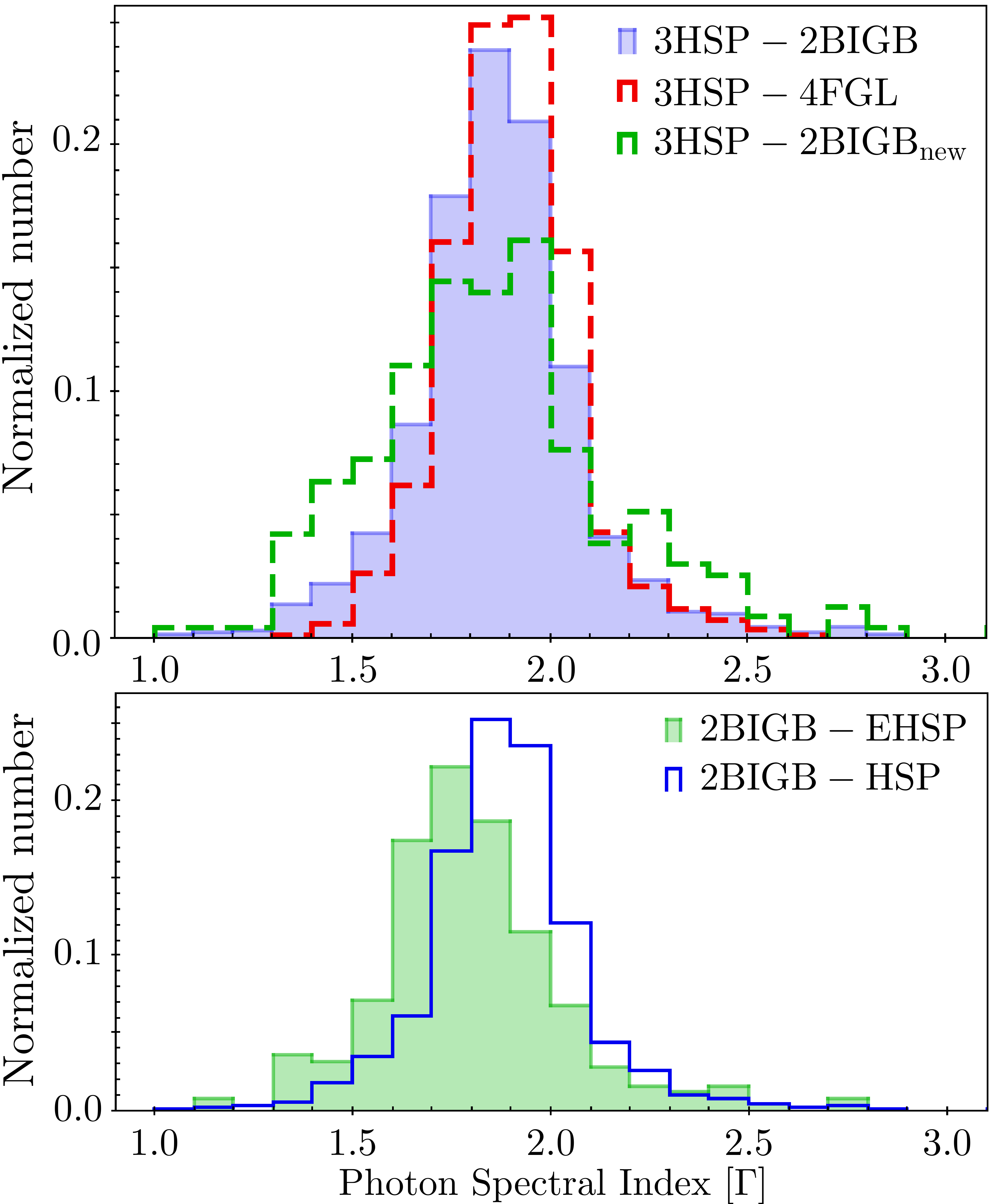}
\caption{(Top) Normalized histogram to compare the photon index $\Gamma$ distribution for the samples 3HSP-2BIGB (full indigo bars), 3HSP-4FGL (red dashed), and 2BIGB$_{new}$ (green dashed). In the case of 3HSP-4FGL sources, the $\Gamma$ values correspond to 4FGLv20. (Bottom) Normalized histogram to compare $\Gamma$ for 2BIGB sources divided into HSP and EHSP subsamples, respectively, with 908 and 252 objects.}
\label{fig:slope-histo}
\end{figure}

% 2BIGB Total: 1160 sources; 925 in 4FGL + 235 new  (9 were reported in previous FGL/FHL cats)
Fig. \ref{fig:slope-histo} (top) shows the photon spectral index distribution for all the 925 3HSP-4FGL and compare it to the 1160 3HSP-2BIGB sources. We also plot the distribution for the 235 2BIGB$_{new}$ sources, which are new detections concerning 4FGL. A Kolmogorov-Smirnov (KS) test shows that the 3HSP-4FGL and 3HSP-2BIGB normalized histograms are similar, with a p-value of 0.587. Also, the normalized distribution of 3HSP-4FGL and 3HSP-2BIGB$_{new}$ are similar under the KS test, with a p-value of 0.174. Therefore, at 5\% level (p-value\,>\,0.05), the photon index distributions are alike and consistent with a single-parent population.

The $\gamma$-ray photon index deduced from the 2BIGB analysis agree with 4FGL, for both the entire 2BIGB sample and the 2BIGB$_{new}$ subsample. The mean values $\langle \Gamma \rangle$ and the distribution width ($\pm$1$\rm \sigma$) for each sample are $\langle \Gamma_{\rm 3HSP-2BIGB} \rangle$\,=\,1.87$\pm$0.21, $\langle \Gamma_{\rm 2BIGB-new} \rangle$\,=\,1.86$\pm$0.32, and $\langle \Gamma_{\rm 3HSP-4FGL} \rangle$\,=\,1.90$\pm$0.17. Moreover, the $\langle \Gamma_{\rm 3HSP-4FGL} \rangle$ parameter (as derived from the associations done in this work, Table \ref{table:2BIGB-10yrs-slim}) is in full agreement with the value obtained in 4LAC \citep{4lac} for the HSP population, $\langle \Gamma_{\rm HSP-4LAC} \rangle$\,=\,1.90$\pm$0.17. 

%[*]Add the mean values. 
% Gamma_4FGL      = 1.9042 \pm 0.1681/sqrt(925)  (=0.005)
% Gamma_2BIGB     = 1.8682 \pm 0.2180/sqrt(1160) (=0.006)
% Gamma_2BIGB_new = 1.8549 \pm 0.3192/sqrt(235)  (=0.02 )

In Fig. \ref{fig:slope-histo} (bottom), we divide the 2BIGB catalogue into subsamples of HSPs (908 objects) and EHSPs (252 objects) to compare their photon index $\Gamma \rm _{0.5-500\,GeV}$ histograms. A KS test to compare the normalized histograms shows that the distributions are alike (p-value of 0.838), and with compatible mean values ($\pm$1$\rm \sigma$) of $\langle \Gamma_{\rm 2BIGB-HSP} \rangle$\,=\,1.89$\pm$0.20 and $\langle \Gamma_{\rm 2BIGB-EHSP} \rangle$\,=\,1.80$\pm$0.25. The $\langle \Gamma \rangle$ values measured for the HSP and EHSP subsamples are similar, in agreement with a flattening in the log($\rm \nu_{syn-peak}$) vs. Photon Index plane, which is observed in 4LAC \citep{4lac} for sources with synchrotron peak $\rm \nu_{syn-peak}$>10$^{15}$\,Hz. 

\begin{figure}
\includegraphics[width=1.0\linewidth]{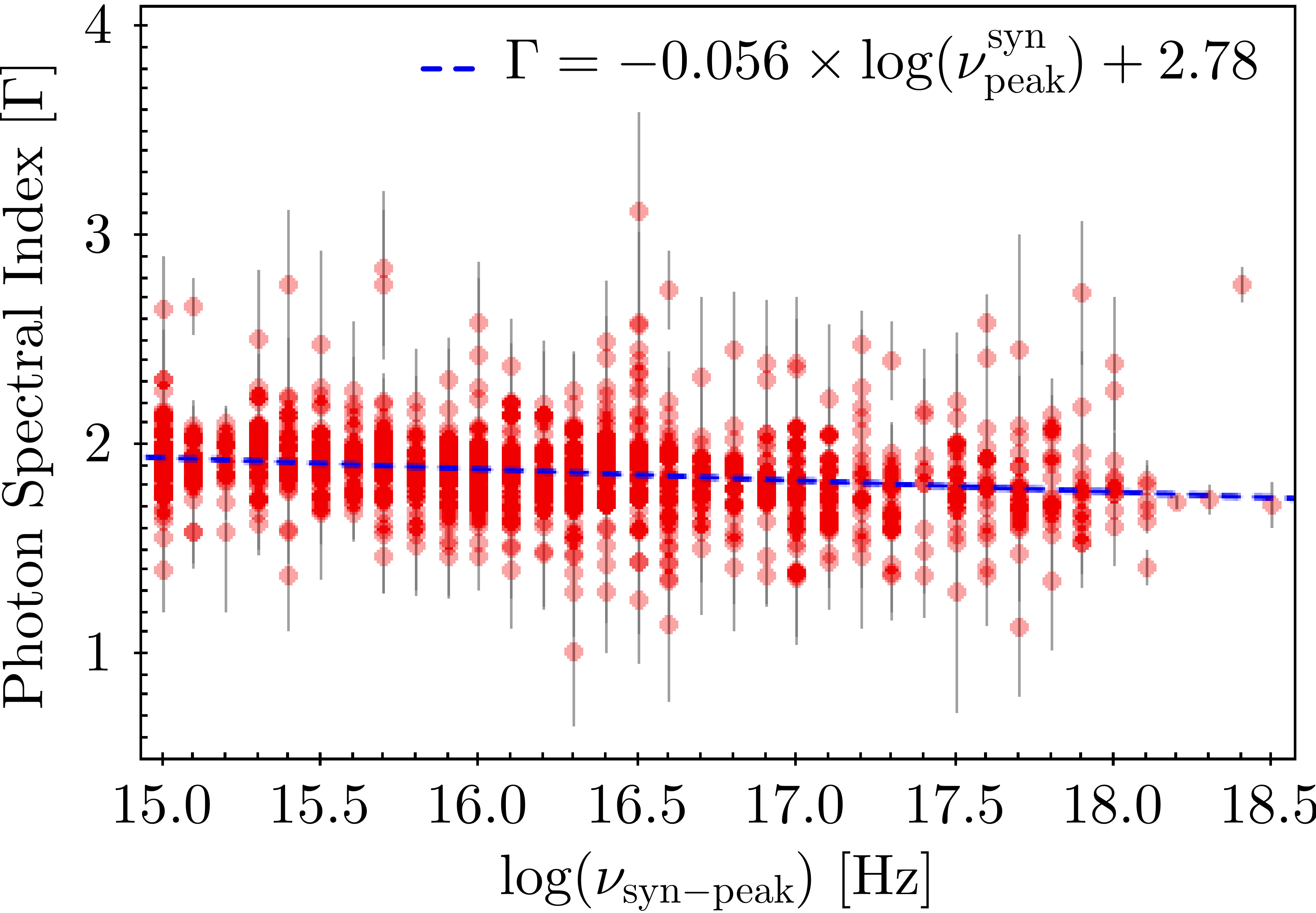}
\caption{Scatter plot with the 0.5-500\,GeV photon spectral index $\Gamma$ versus the log of synchrotron peak log($\rm \nu_{syn-peak}$) for the 3HSP-2BIGB sample. The blue dashed line is a linear fit to the data.}
\label{fig:slope-vs-nupeak}
\end{figure}

Fig. \ref{fig:slope-vs-nupeak} shows the Photon Index vs. log($\rm \nu_{syn-peak}$) relation for the 3HSP-2BIGB sample and confirm the flattening at $\rm \nu_{syn-peak}$>10$^{15}$\,Hz. We find a relatively week trend of hardening $\Gamma$\,=\,-0.056$\times$log($\rm \nu_{syn-peak}$)+2.78, with increasing synchrotron peak frequency. However, according to 4LAC paper \cite{4lac}, the hardening trend is stronger when considering the entire blazar population, with $\rm \langle \Gamma \rangle$\,$\sim$2.6 at log($\rm \nu_{syn-peak}$)=12.0[Hz], and $\rm \langle \Gamma \rangle$\,$\sim$1.9 at log($\rm \nu_{syn-peak}$)=16.0[Hz].

We note that there is no relevant correlation between redshift and photon spectral index for the 2BIGB sample alone\footnote{The 4LAC paper \citep{4lac} also reports on the absent (or low significance) correlation between redshift and $\rm \Gamma$ when considering blazar subclasses alone.}. Nevertheless, when considering all blazar families as in 4LAC paper, this correlation shows up as an average $\gamma$-ray spectral softness for sources with large redshift. The 4LAC attributes this effect to the intrinsic spectral curvature at the high energy end, which is redshifted to lower energies as z increases, and result in softer spectrum. However, a significant fraction of the $\rm \gamma$-ray detected BL Lacs have no spectroscopic redshift measurements. Taking the 2BIGB catalogue as an example, only 377 out of 1160 3HSP-2BIGB sources (32.5\%) have a reliable redshift measurement (redshift flag-1 in Table \ref{table:2BIGB-10yrs-slim}). Therefore any correlation with redshift or derived quantities, like source luminosity, could be under heavy bias. 

\begin{figure}
\includegraphics[width=1.0\linewidth]{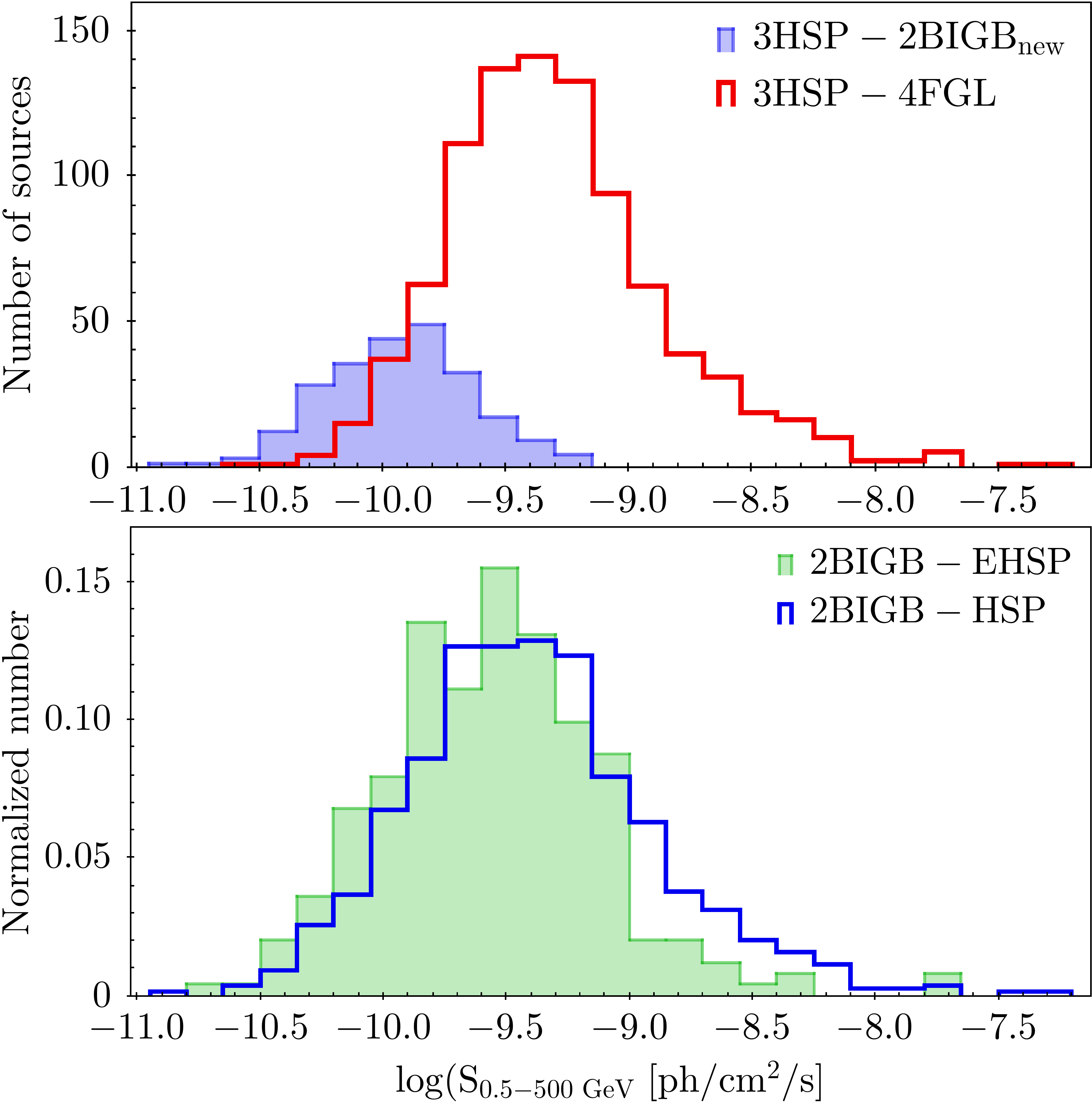}
\caption{(Top) Histogram to compare the log of integral $\rm \gamma$-ray flux log(S$\rm _{0.5-500 \ GeV}$) for the 3HSP-4FGL (red line) and 2BIGB$_{new}$ (full indigo bars). (Bottom) Histogram to compare the integral $\rm \gamma$-ray flux for 2BIGB sources divided in HSP (blue line) and EHSP (full green bars) subsamples.}
\label{fig:flux-histo}
\end{figure}

The Fig. \ref{fig:flux-histo} (top) shows the integral flux distribution S$\rm _{(0.5-500\,GeV)}$ for the 3HSP-4FGL sample and to compare with the 2BIGB$_{new}$, from where it is seen that the 2BIGB$_{new}$ detections represent an underlying population of faint $\rm \gamma$-ray emitters. Fig. \ref{fig:flux-histo} (bottom) shows the S$\rm _{(0.5-500\,GeV)}$ distribution, dividing the entire 2BIGB catalogue into HSP and EHSP subsamples, with 908 and 252 objects, respectively. A KS test to compare the normalized histograms confirms the distributions are similar (p-value\,=\,0.681) and shows that EHSPs are not necessarily fainter than HSPs concerning the photon-counts in the {\it Fermi}-LAT band. 

Also, for the 3HSP-2BIGB sample, we study the synchrotron peak versus the gamma-ray flux plane, $\rm \nu_{syn-peak}$ vs. S$\rm _{(0.5-500\,GeV)}$, and report in null correlation (or very weak), with a Pearson correlation coefficient of $\sim$\,-0.2. Given the absent correlation between synchrotron $\rm \nu_{syn-peak}$ with S$\rm _{(0.5-500\,GeV)}$ and the flattening concerning $\Gamma_{\rm 0.5-500 \, GeV}$, we find that the HE spectral characteristics of HSPs and EHSPs are similar. Therefore, the search for TeV peaked blazars (or Extreme TeV BL Lacs) as done in \cite{Mini-ETeV-Cat-2019} may benefit from considering both HSP \& EHSP blazars as TeV candidates, instead of EHSPs only. In agreement to that,  \cite{TeV-peak-candidates-Costamante2019} who searches for TeV peaked blazars considering spectral properties of the entire 5BZcat and Sedentary catalogues \citep{5BZcat-V5,SedentaryI}, and delivers a list with 47 Extreme TeV BL Lacs candidates.    

We call attention to the argument that ``EHSPs might peak at E$>$1\,TeV while the HSP population is more likely to peak at the energy window covered by {\it Fermi}-LAT'', which is not precise. As reported in \cite{EHSP-sample-Foffano2019}, the HE properties of TeV detected EHSP sources are very similar but can be substantially different at VHE. The VHE spectrum of EHSP sources comprises a mix of cases peaking at hundreds of GeV and cases where it is possible to observe a hard \& continuum spectrum from HE to VHE\footnote{All VHE spectra considered in \cite{EHSP-sample-Foffano2019} are deabsorbed with the EBL model from \cite{EBL-model-Franceschini2017}.}.

\subsection{The contribution of HSP \& EHSP blazars to the EGB}

\begin{figure}
\includegraphics[width=1.0\linewidth]{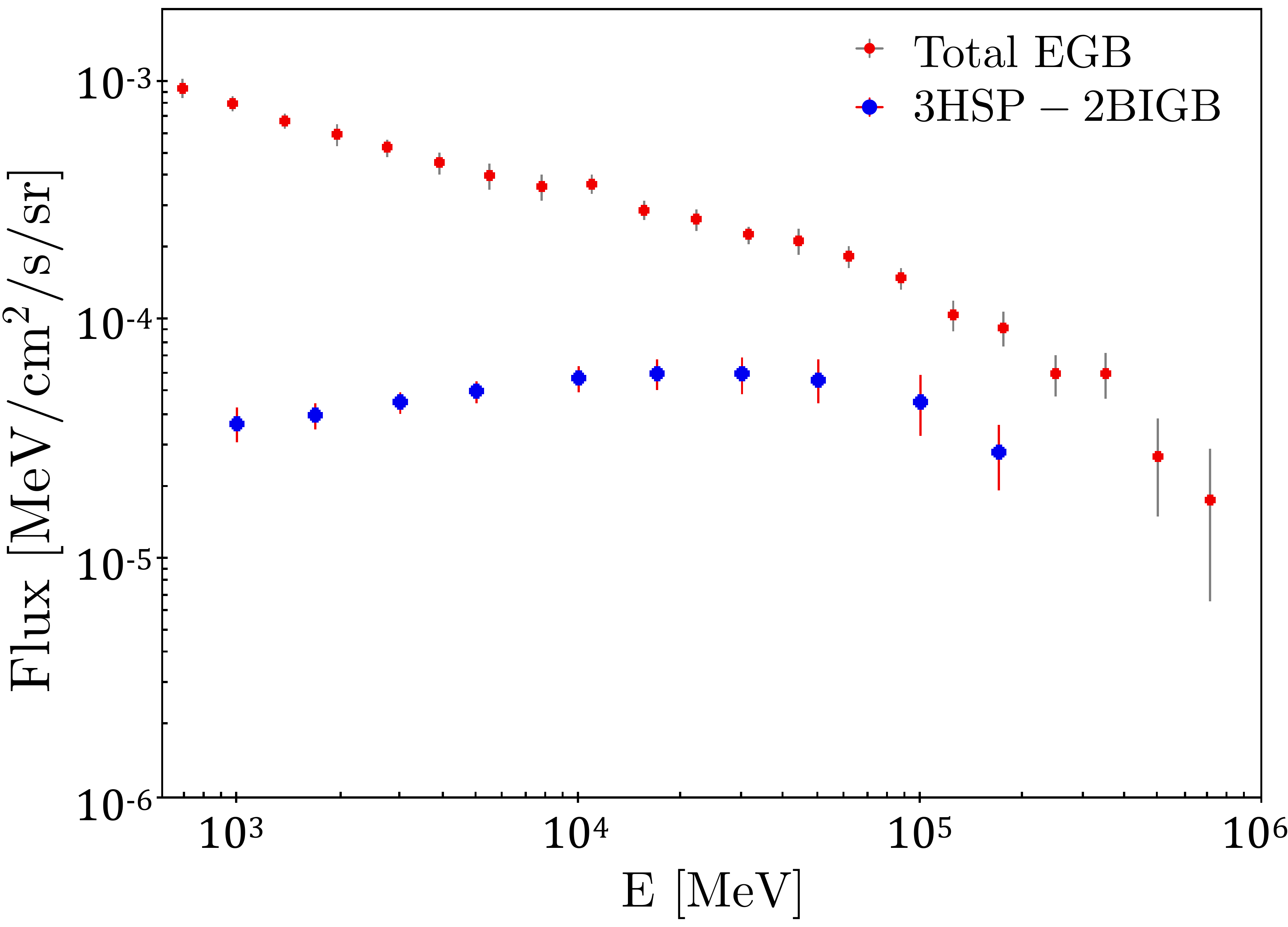}
\caption{The contribution of 3HSP-2BIGB sources to the EGB. The blue points represent the stacked $\rm \gamma$-ray SED for the entire 3HSP-2BIGB population at |b|>10$^\circ$, and red for the total extragalactic gamma-ray content \citep[EGB,][]{Ackermann2015-EGB}. The signals are averaged over the sky area at |b|\,>\,10$^\circ$, and |b|\,>\,20$^\circ$, respectively.}
\label{fig:EGB}
\end{figure}

Here we investigate the total - measured - contribution of HSP \& EHSP blazars to the extragalactic gamma-ray background. We build a stacked SED with the high galactic latitude 2BIGBs (|b|>10$^\circ$) as a sum of fluxes for each bin E$\rm _0$ (all UL fluxes are removed from the sum), and the stacked flux for each energy E$_0$ is averaged over the sky area  A$_{\rm |b|>10^\circ}$\,=\,34,110.3 deg$^2$\,=\,10.39 sr. Fig. \ref{fig:EGB}, shows the integral spectral contribution of HSP + EHSP blazars to the total extragalactic $\rm \gamma$-ray content (EGB). The total EGB flux is measured for high galactic latitude (|b|>20$^\circ$) and considers the foreground model A from \cite{Ackermann2015-EGB,IsoDiffuseGammaBackGround}. Note that Fig. \ref{fig:EGB} represents a direct and unprecedented measurement of the emission produced by resolved HSP \& EHSP blazars, down to the faintest $\gamma$-ray signatures at the TS\,>\,9 level. 

% FIGURE FROM NEXT SECTION

Table \ref{table:EGB} shows the 2BIGB improvement regards to solving the extragalactic component into point sources. We add 235 new $\gamma$-ray emitters to the known population of 925 3HSP-4FGL sources, which brings a relatively small -but not negligible- contribution to the total intensity already solved by 4FGL. The 2BIGB represents an improvement of $\sim$25\% for the number of 3HSP blazars detected in $\gamma$-rays. However, the entire group of 2BIGB$_{new}$ sources only populate the faint end of the log(S$\rm _{0.5-50 \, GeV}$) flux distribution (see Fig. \ref{fig:flux-histo}), meaning that the most intense 3HSP-4FGL sources already dominate the stacked fluxes. The extra contribution solved into 2BIGB$_{new}$ sources ranges from 1.2\% up to 3.4\% along the 1\,GeV to 170\,GeV band (column `I(\%)' in Table \ref{table:EGB}) peaking around the 17-30\,GeV energy band. 

There is, in fact, an underlying population of faint gamma-ray sources that contribute to the EGB with a spectral index characteristic of HSP blazars. Moreover, the total contribution of HSP \& EHSP blazars to the EGB gets more relevant with increasing energy (column `Tot(\%)', Table \ref{table:EGB}), reaching up to 33.5\% at 100\,GeV. This trend is in agreement with predictions from \citep{giommisimplified2,EGBpaolo,DiffuseGammaBLlacs}, which argue that HSP blazars are the dominant fraction of isotropic $\rm \gamma$-ray background at the E\,>\,10\,GeV. Besides, \cite{EGB-Blazar-Contribution-DiMauro2018} concludes that the blazar population may represent up to (42$\pm$8)\% of the total EGB at E\,>\,10\,GeV. 

The measured fraction of the EGB due to HSP blazars, drop from 33.5\% -at 100\,GeV- to 29.9\% -at 170\,GeV- in what could be the result of lower detection efficiency at the highest energy bands. Also, we adopt a more restrictive upper limit threshold (TS\,<\,6 $\rightarrow$ TS\,<\,10) for the 100\,GeV and 170\,GeV bins. Even though knowing this would affect the detection efficiency at those bands, this is done to assure the VHE detections are robust. For the same reasons, the flux fraction associated with 2BIGB$_{new}$ sources (Flux$\rm _{2BIGB-new}$\,/\,Flux$\rm _{2BIGB}$, column `I(\%)' in Table \ref{table:EGB}) reaches 3.3\% at 50\,GeV, but also see a drop for the largest energy bins of 100\,GeV and 170\,GeV. There are indeed significant uncertainties at VHE for both the EGB and the 2BIGB integral fluxes. Notably, at 170\,GeV, the errors are enough to absorb the observed drop in `Tot(\%)' flux-ratio (Flux$\rm _{2BIGB}$\,/\,Flux$\rm _{EGB}$) as a fluctuation.

\begin{table}
\centering
\caption{The total contribution of HSP \& EHSP blazars to the EGB content as measured from the 3HSP-2BIGB and 3HSP-4FGL samples. Flux in [MeV/cm$^2$/s/sr], for the Total EGB, and for the resolved contribution from 3HSP-2BIGB and 3HSP-4FGL sources. The `I(\%)' column shows the relative improvement of the 2BIGB concerning 4FGL for solving the EGB, and the `Tot' column list the total fraction (\%) of the EGB solved into HSP \& EHSP blazars given the 2BIGB data.}
\label{table:EGB}
{\def\arraystretch{1.4}
\begin{tabular}{lccccc}
\hline
E$_{\rm GeV}$ & Flux$\rm ^{EGB}_{(\times10^{-4})}$ & Flux$\rm ^{4FGL}_{(\times10^{-5})}$ & Flux$\rm ^{2BIGB}_{(\times10^{-5})}$ & I(\%) & Tot(\%)  \\
\hline
 1.0  & 8.11$\pm$0.52  &  3.56$\pm$0.53 &  3.65$\pm$0.57  & 2.71 & 4.5$^{\plus 1.0}_{-0.93}$  \\      
 1.7  & 6.30$\pm$0.46  &  3.88$\pm$0.42 &  3.99$\pm$0.47  & 2.78 & 6.3$^{\plus1.3}_{-1.1}$    \\      
 3.0  & 5.13$\pm$0.39  &  4.38$\pm$0.38 &  4.50$\pm$0.44  & 2.98 & 8.7$^{\plus1.6}_{-1.4}$    \\      
 5.0  & 4.16$\pm$0.45  &  4.84$\pm$0.42 &  4.99$\pm$0.48  & 3.14 & 11.9$^{\plus2.7}_{-2.2}$   \\      
 10   & 3.66$\pm$0.34  &  5.51$\pm$0.58 &  5.69$\pm$0.66  & 3.23 & 15.5$^{\plus3.6}_{-3.0}$   \\      
 17   & 2.83$\pm$0.25  &  5.79$\pm$0.69 &  5.99$\pm$0.78  & 3.44 & 21.1$^{\plus5.1}_{-4.2}$   \\      
 30   & 2.33$\pm$0.17  &  5.78$\pm$0.91 &  5.98$\pm$1.01  & 3.43 & 25.6$^{\plus6.7}_{-5.8}$   \\      
 50   & 2.05$\pm$0.21  &  5.43$\pm$0.10 &  5.61$\pm$1.10  & 3.33 & 27.3$^{\plus9.1}_{-7.4}$   \\      
 100  & 1.34$\pm$0.14  &  4.36$\pm$0.12 &  4.49$\pm$1.24  & 2.84 & 33.5$^{\plus14.2}_{-11.5}$ \\      
 170  & 0.937$\pm$0.14 &  2.74$\pm$0.79 &  2.78$\pm$0.82  & 1.23 & 29.9$^{\plus15.7}_{-11.6}$ \\  
\hline
\end{tabular}}
\end{table}

Note that the flux at each pivot-energy is calculated over the E-bins listed in Table \ref{table:E-bin}. The fit for each E-bin had both Normalization N$_0$ and Photon Index $\Gamma$ free to vary. Note, the SED analysis is not required to be bound to the broadband fit over 0.5-500\,GeV, and therefore it is more sensitive to the spectrum curvature at the highest energy bands. Alternatively, if one uses the power-law broadband fit (0.5-500\,GeV) to estimate fluxes at the high-energy bands, the risk is to be overestimated. 

The power-law broadband fit is dominated by the lower energy photons \citep{4lac}, which have larger counts-rate and tend to determine the broadband photon index at a regime where the absorption due to EBL is not relevant. As a result, the power-law broadband fit tends to overestimate the VHE flux since the spectrum curvature is not described properly. Our approach avoids this bias, given that the SED data-points are computed based on multiple energy-bands that adjust well to the observed spectrum curvature. Therefore the SEDs provide a confident way to measure the integral contribution of point-sources to the very high-energy EGB.

Note that our work does not rely on extrapolations of the $\gamma$-ray SED or correlations to other wavebands \citep[as in,][]{EGB-from-Radio-extrapolation-DiMauro2013a,EGB-from-Radio-extrapolation-DiMauro2014}, neither detection efficiency corrections \citep[as in,][]{EGB-Blazar-Contribution-DiMauro2018}. An alternative and robust approach based on stacking analysis is discussed in \cite{Stack-Gamma-Pop-Paliya2019}, and has similarities to our work, given that multifrequency selected seeds drive the search for the stacked signature. 

Accurate measurement of the EGB content produced by $\rm \gamma$-ray resolved blazars can be accomplished by large-scale spectrum analysis (as done here), looking for signatures down to a low detection threshold and for the entire blazar population. The advantage is to produce detailed descriptions for individual sources while solving the EGB content. This allows to cross-check the measured flux at each energy bin with the estimates and extrapolations from other works, which should converge.

\subsection{The gamma-ray logN-logS of 2BIGB sources}
\label{section:pop}

The number counts plot (logN-logS) is a valuable tool to visualize and test if a population of astrophysical sources is uniformly distributed at low redshift. If the cumulative distribution follows the so-called Euclidean trend $\propto$S$^{-1.5}$, it is a strong indicative the sample is complete and non-evolving at least down to the flux-limit where it detaches from that trend. The measured number count is expected to deviate from the Euclidean slope at lower fluxes because of cosmological effects. However, it can also detach at intermediary fluxes as a result of the luminosity evolution of the population along time \citep{logN-logS-Shanks1984}. 

The logN-logS interpretation can suffer from strong bias introduced by sample incompleteness, given that faint sources can be overwhelmingly numerous and hard to detect or select in its totality. Sample incompleteness can originate from multiple instrumental and source-selection limitations, and a clear understanding of those inefficiencies is crucial to derive conclusions from the logN-logS shape. To understand how a population evolves along with cosmic history, one needs to rely on a complete sample with robust redshift information. Though, a significant fraction of HSP blazars (68\%, 1373 out of 2013 3HSPs) has no spectroscopic redshift, which hinders the understanding of blazar's evolution. 

For blazars, it is not clear where to expect deviations from the Euclidean trend in the number counts, and here we look into that. The observed $\gamma$-ray logN-logS associated to 3HSP sources is shown, and also for the HSP and EHSP subsamples alone. The main features observed are discussed in details, and future works  could apply improvements, e.g., to consider the effects of detection efficiency and k-correction.    

\begin{figure}
\includegraphics[width=0.99\linewidth]{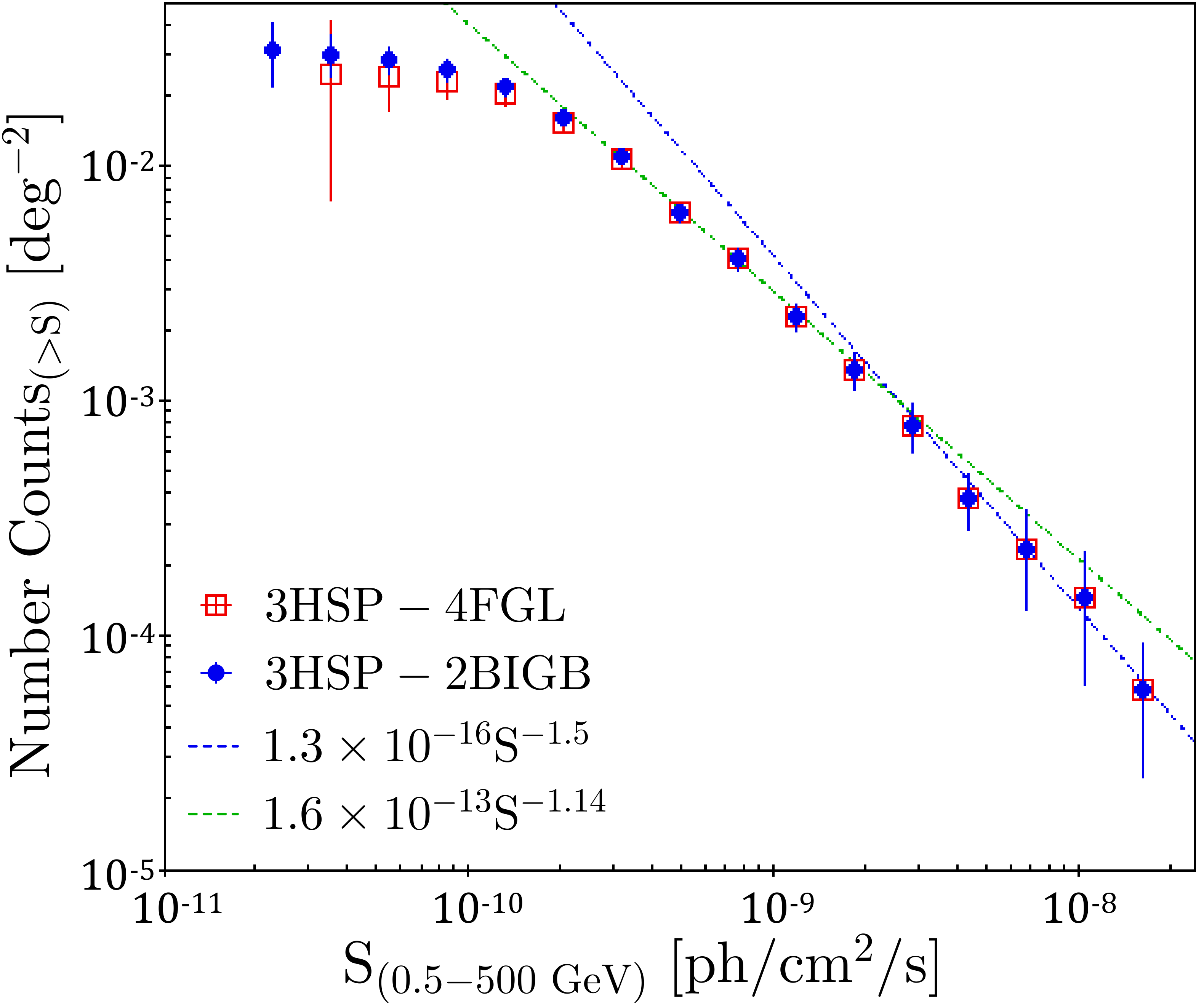}
\caption{The $\rm \gamma$-ray logN-logS for the 3HSP population detected in 2BIGB (blue) and in 4FGL (red), plotting the cumulative density of sources N with flux lager than the integral flux S$ \rm _{0.5-500 \ GeV}$, only considering high galactic latitude sources at |b|>10$^\circ$.}
\label{fig:logNS}
\end{figure}

Fig. \ref{fig:logNS} shows the $\gamma$-ray logN-logS for the 3HSP population detected in 2BIGB and in 4FGL, and only consider sources at |b|>10$^\circ$ where the 3HSP selection is homogeneous. This encloses a sky-area of 34110.3 deg$^2$, which holds the majority of 3HSP sources (1925 out of 2013), from where 1073 are 3HSP-2BIGB, and 840 are 3HSP-4FGL. 

As in previous work \citep{1BIGB}, we observe an early-break in the S$^{-1.5}$ Euclidean trend at $\sim$2.5$\times$10$^{-9}$. One could argue that this might be a bias because of the {\it Fermi}-LAT exposure, which is not constant along the entire sky. If it were the case, our analysis would have seen improvements concerning completeness at the level of $\rm S_{(0.5-500\,GeV)}$\,<\,2.5$\times$10$^{-9}$ ph/cm$^2$/s, since the 2BIGB catalogue considers extra three years of observations when compared to 4FGL. 

\begin{figure}
\includegraphics[width=1.0\linewidth]{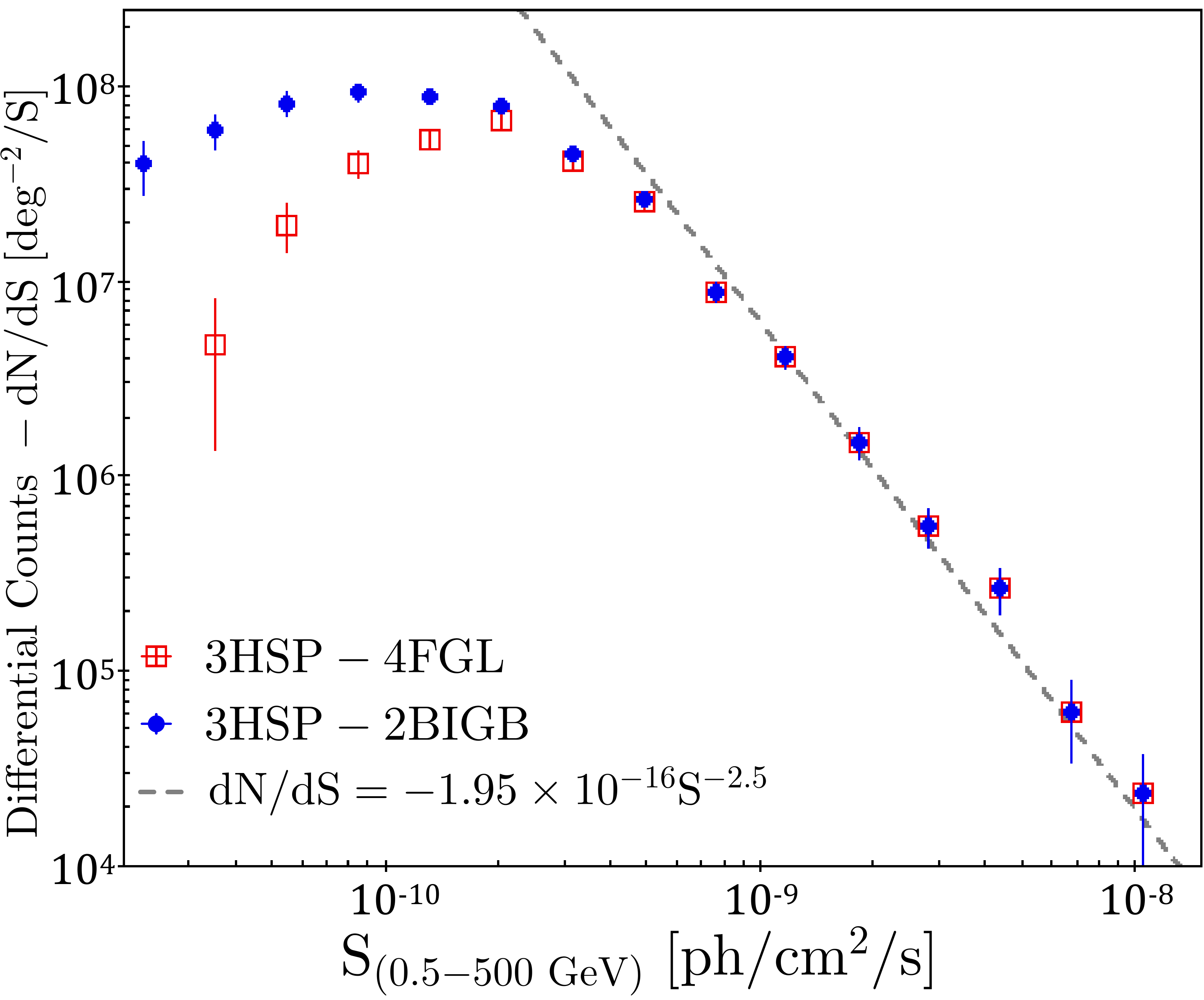}
\caption{The differential number counts (-dN/dS) in $\rm \gamma$-ray for the 3HSP population, comparing the 3HSP-2BIGB (blue) and 3HSP-4FGL (red) samples. The dashed line is the derivative of N\,=\,1.3$\times$10$^{-16}$S$^{-1.5}$, which is the Euclidean fit to the number counts from Fig. \ref{fig:logNS}.}
\label{fig:dnds}
\end{figure}

The 1BIGB paper reported this same early-break feature, which also highlights the existence of previous logN-logS data from the 3LAC paper \citep{3lac} that has similar characteristics as in here (see footnote 19 at \cite{1BIGB}). The existence of an early break in the $\gamma$-ray logN-logS plane is intriguing, and its origin is not clear. However, it could well be the result of $\gamma$-ray variability, incompleteness from the 3HSP catalogue itself, absorption, or the need to apply k-correction to the observed flux. Therefore, proper treatment needs to add corrections to the $\gamma$-ray logN-logS, which should impact further discussions related to source evolution (see \cite{50GeVbackground} as an example). To mention in more details: 

\begin{itemize}
    \item Gamma-ray flares. When integrated along 11 years of observations, a flare event in $\gamma$-rays gets diluted, which overestimates the mean flux concerning the non-flaring state of the source. This bias could act to exaggerate the number of sources in the bright end of $\gamma$-ray logN-logS.
    \item 3HSP incompleteness. As mention in 1BIGB paper (for the case of 2WHSP catalogue), the early-break could be a manifestation of incompleteness from the 3HSP catalogue that increasingly affects the selection of fainter sources as their synchrotron peak flux gets lower. Currently available radio surveys (SUMMS and NVSS) and the poor all-sky coverage in X-rays could be the actual limitations. The 3HSP paper reports on blazars with yet undetected radio counterpart, most likely with a faint radio-flux, which is lower than the flux-limit from SUMMS, NVSS, and FIRST radio surveys \citep{3HSP}. An estimate for the number of similar cases is still to be evaluated, and could represent a significant fraction of the HSP \& EHSP population.     
    \item Absorption. This is a $\gamma \gamma$ process where a VHE photon in a head-on collision to extragalactic background light (EBL) annihilates to create electron-positron pairs. Absorption affects high redshift sources more heavily and steepens the observed spectrum, depleting the S$\rm _{(0.5-500 \ GeV)}$ flux. As highlighted by \cite{EGB-Blazars-Costamante2013}, the critical energy above which at least 5\% of the emitted photons are absorbed due to EBL is given by E$\rm _{crit \left( z \right)}$\,=\,170\,(1+z)$^{-2.38}$ \,GeV \citep{EBL-imprint-on-blazars-FermiLAT2012}, based on EBL model from \cite{EBL-model-Franceschini2008}. Given that the 3HSP catalogue has confident lower-limit redshifts up to 0.7, the $\rm E_{crit \left( z \right)}$ can be as low as 48\,GeV. 
    \item {\it Fermi}-LAT exposure. The {\it Fermi}-LAT data taken mode is turned off during passages along the South Atlantic Anomaly, inducing up to 15\% sensitivity difference between the north and south hemisphere. As mention previously, this does not seem to be the main driver for the early-break in the $\gamma$-ray logN-logS. 
    \item K-correction. To properly apply k-correction to the observed $\gamma$-ray flux, a redshift survey for the entire 3HSP catalogue is necessary. Indeed, the lack of spectroscopic (robust) redshift measurements for $\sim$68\% of the 3HSPs is a limitation. The use of photometric redshifts can help to close that gap, although introducing a certain degree of uncertainty. The 3HSP catalogue has estimated photometric redshifts for 930 objects, but still, nearly 1/4 of the sources lack a redshift estimate. One way to apply the correction is to compute the intrinsic gamma-ray flux S$ _{E'_i-E'_f}$ given that the observed energy interval (E$_i$ to E$_f$, selected to integrate over) gets corrected based on redshift according to E$_i$=E$'_i$/(1+z) and E$_f$=E$'_f$/(1+z). This way, one can probe the same intrinsic energy range, from E$'_i$ to E$'_f$. The selection of a suitable energy range can minimize the effect of absorption due to EBL.  
\end{itemize}

\begin{figure}
\includegraphics[width=1.0\linewidth]{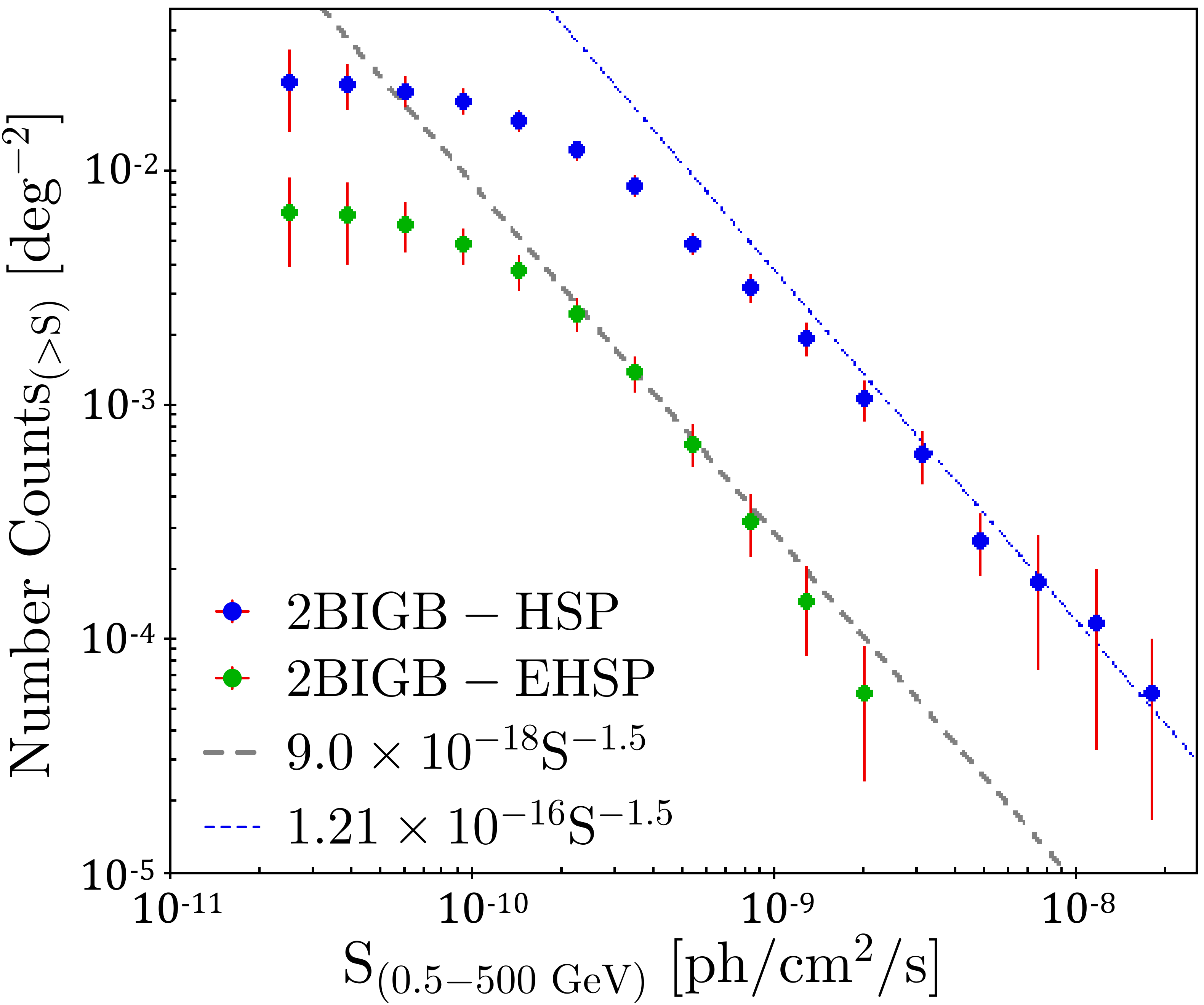}
\caption{The $\rm \gamma$-ray logN-logS for the entire 3HSP-2BIGB population, considering the integral flux S$ \rm _{0.5-500 \ GeV}$ and separating between the HSP (blue) and EHSP (green) subsamples. Blue and gray dashed lines are fits to the HSP and EHSP logN-logS respectively, following the S$^{-1.5}$ slope for a non-evolving population.}
\label{fig:logNS-HSP-EHSP}
\end{figure} 
 
In Fig. \ref{fig:dnds}, the differential counts dN/dS vs. S is shown, and a fast drop for S\,<\,2\,$\times$10$^{-10}$ ph/cm$^2$/s is seen. This drop is most likely because the detection efficiency decreases with flux, meaning that faint sources are harder to detect. Moreover, when comparing the 3HSP-4FGL to the 3HSP-2BIGB sample, there are gains concerning completeness for fluxes close to $\sim$2$\times$10$^{-10}$ ph/cm$^2$/s and lower. This is because of the larger exposure from 2BIGB (11 years) regarding 4FGL (8 years). 

Apart from the early-break feature, the $\gamma$-ray sample seems complete down to the flux level of $\sim$2$\times$10$^{-10}$ ph/cm$^2$/s. From there on to lower fluxes, the 2BIGB catalogue introduce improvements in the differential number counts compared to 4FGL. The 2BIGB is indeed more complete than 4FGL in the lower flux end, as expected from Fig. \ref{fig:flux-histo}.

Considering that the 2BIGB catalogue is currently the most sensitive $\gamma$-ray survey over the HSP \& EHSP populations, it is also suitable to study the sky-density of each blazar class. Fig. \ref{fig:logNS-HSP-EHSP} shows the number counts for the 2BIGB catalogue dividing between HSP and EHSP blazars. Here we consider only the 3HSPs at high galactic latitude |b|>10$^\circ$, with a total of 838 2BIGB-HSP and 235 2BIGB-EHSP objects. 

From Fig. \ref{fig:logNS-HSP-EHSP} we conclude that the density of $\gamma$-ray detected HSPs is $\sim$\,13.4$\times$ larger compared to EHSPs, and that the incompleteness of the EHSP $\gamma$-ray sample starts at $\sim$2$\times$10$^{-10}$ ph/cm$^2$/s; Apart from the early-break feature, this is similar to what is seen in Fig. \ref{fig:dnds} for the entire 2BIGB catalogue. Also, the HE number counts of HSP and EHSP sources -at the bright end- are compatible to the Euclidean trend, indicating that both classes are homogeneously distributed in space.

Concerning the $\gamma$-ray detectability of HSP vs. EHSP sources, the fraction $f$ of each class within the 4FGL and 2BIGB catalogues is very similar, with ($f \rm _{HSP}$ \& $f \rm _{EHSP}$) of (0.789 \& 0.209) for 4FGL, and (0.781 \& 0.219) for 2BIGB. Therefore, we expect that observational improvements in $\gamma$-ray sensitivity and exposure time may unveil new gamma-ray blazars independent of its blazar class. 

%Numbers at |b|>10
%2BIGB: 1073      2BIGB-HSP=838        2BIGB-EHSP=235
%4FGL : 840        4FGL-HSP=663         4FGL-EHSP=176      

\section{Summary and Conclusions}

% 2BIGB Total: 1160 sources; 925 in 4FGL + 235 new  
The 2BIGB catalogue has 1160 sources and is the result of a $\gamma$-ray analysis of the entire 3HSP catalogue. The position of 3HSP sources are used as multifrequency seeds where we test for the existence of a $\gamma$-ray signature. Such an approach has a low number of free parameters, allowing to lower the detection threshold down to TS\,>\,9 without compromising the final catalogue. The likelihood analysis fits only two free parameters (Normalization and Photon index) given the source position is set as fixed. 

All 2BIGB detections result from a broadband 0.5-500\,GeV analysis integrating over 11 years of observations with {\it Fermi}-LAT. We confirm and update the fitting parameters for the 925 2BIGB sources that have a counterpart in 4FGL. Also, the 2BIGB catalogue includes 235 sources that are new detections concerning 4FGL (226 of them were never reported in previous {\it Fermi}-LAT catalogues).  

The fact that 925 2BIGB sources are already part of the 4FGL catalogue allowed us to compare results and to validate our analysis, supporting the robustness of the new $\gamma$-ray detections presented in this work. Moreover, all 2BIBG$_{new}$ were evaluated with high-energy TS maps and confirmed as point sources.

The $\gamma$-ray photon index distribution of 3HSP-4FGL, 3HSP-2BIGB, and 2BIGB$_{new}$ sources, are compared via KS test and show to be compatible with a single parent population. In particular, the 235 new detections have spectral properties that are well representative of the HSP population, excluding the possibility of heavy contamination for the 2BIGB$_{new}$ sample.

We have built the $\gamma$-ray spectral energy distribution for all 2BIGB sources, covering the 1-170GeV energy range with 10.5 years of observations with {\it Fermi}-LAT, and integrating over superposed energy bins. The SED data-points are available at \url{https://github.com/BrunoArsioli} with a template to upload at the ASDC-ASI portal and visualize with the SED Builder Tool \url{https://tools.ssdc.asi.it/}. The data will soon be available at OpenUniverse \url{http://www.openuniverse.asi.it/} and BSDC \url{http://bsdc.icranet.org/} portals\footnote{Find more information about the Open Universe Initiative at \cite{OpenUniverse-2019}, and about the Brazilian Science Data Center BSDC at \cite{BSDC-2017}.}. 

This work revealed an underlying population of $\gamma$-ray emitters at the threshold detectability for {\it Fermi}-LAT. Those sources are not necessarily of low relevance and contribute to the EGB with photon index characteristic of HSP blazars, however, with fainter flux. Many of them have a hard photon spectral index, and only seen at the largest energy bands. Those cases can be relevant at higher energies and likely at reach for the upcoming Cherenkov Telescope Array (CTA). Therefore, this work delivers a complementary description of the $\gamma$-ray sky, which could impact on VHE population studies for CTA. 

Regarding the $\gamma$-ray background, we show that the measured contribution of 3HSP blazars (HSP+EHSP) to the total EGB increases with energy, and reach 33.5\% at 100\,GeV. The 235 new detections serve to solve a fraction of the EGB into point sources, shrinking the available space for an actual extragalactic diffuse $\gamma$-ray component. When considering the additional 2BIGB$_{new}$ sources, the resolved fraction of the EGB flux at 50 GeV is improved by 3.3\%. The enhancements for other energy bands are lower, but of the same order, and listed in Table \ref{table:EGB}.

We plot the $\gamma$-ray number counts (logN-logS), and also, the differential number counts for the 3HSP-2BIGB sources, and compare to the 3HSP-4FGL sample. It is shown that the $\gamma$-ray sample is likely complete down to S$\rm _{(0.5-500 \ GeV)} \, \sim$\,2$\times$10$^{-10}$ ph/cm$^2$/s. We report on an early break in the logN-logS number counts at $\sim$\,2.5$\times$10$^{-9}$ ph/cm$^2$/s, which can not be associated with incompleteness of the $\gamma$-ray sample. This could be a bias arising from $\gamma$-ray variability, incompleteness from the 3HSP catalogue itself, or the need for a proper k-correction. This particular feature in the $\gamma$-ray number counts for HSPs is present since early works \citep{1BIGB,3lac} and demands more investigation. Also, we study the $\gamma$-ray number counts for the HSP and EHSP subclasses. Their distribution in space is relatively homogeneous (compatible with the Euclidean trend at the bright end), and the density of $\gamma$-ray detected HSPs is $\sim$\,13.4$\times$ larger than EHSPs. 

\cite{EHSP-sample-Foffano2019} shows that EHSPs are not all necessarily TeV peaked blazars, but a mix of sources peaking at few to hundreds of GeV, with cases that have a hard photon index from HE to VHE, with a peak at >1\,TeV to >10\,TeV. For the 2BIGB catalogue, we could compare the distribution of HE flux S$_{\rm 0.5-500\,GeV}$ and photon index $\Gamma$ for the HSP and EHSP subclasses, and found they are similar. Therefore, given the HE spectral similarities between those subclasses, the search for Extreme TeV BL Lac candidates may benefit from considering both HSPs and EHSPs as a whole, as done by \cite{TeV-peak-candidates-Costamante2019}. Alternatively, at least, by lowering the synchrotron frequency threshold to incorporate candidates with $\nu_{syn-peak}$\,<\,10$^{17}$Hz. Methods to select TeV-peaked blazars from the entire 3HSP catalogue (i.e. HSP\,+\,EHSP) can also incorporate considerations about HE variability, since TeV-peaked blazars have shown to be remarkably stable (or long-term) HE \& VHE sources \citep{EHSP-nuStar-view-Costamante2018}. 

We show that a direct search for gamma-ray sources driven by multifrequency selected seeds is indeed fruitful and complementary to the official {\it Fermi}-LAT catalogue releases. An extension of current work is highly motivated, intending to account for the totality of blazars and blazar-candidates.

\section{acknowledgements}
     
During this work, BA was supported by S\~ao Paulo Research Foundation (FAPESP) with grant n. 2017/00517-4. BA would like to thank Prof. Marcelo M. Guzzo for his support to this research proposal. BM thank Uppsala University International Science Programme (ISP) for the financial support towards developing this work.  We thank the Centro de Computa\c{c}\~{a}o John David Rogers (CCJDR) at IFGW Unicamp, Campinas-Brazil, for granting access to the Feynman and Planck Clusters. We  thank IcraNet, Prof. Remo Ruffini and Prof. Carlo Bianco for the cooperation and granted access to Joshua Computer Cluster (Rome-Italy). The availability of computational resources was key to the development of our work. The VO publication of our data (\url{vo.bsdc.icranet.org}) is made by the Brazilian Science Data Center (BSDC) service maintained at CBPF, Rio de Janeiro, and accessible through the United Nations Open Universe Initiative at http://www.openuniverse.asi.it. We thank the entire {\it Fermi}-LAT collaboration for manteining a public mission database, which promotes discoveries involving the entire $\rm \gamma$-ray community in a multitude of scientific efforts. We thank J. Biteau for his help in obtaining VHE data from published papers which is now available via OpenUniverse and SSDC-ASI Data Science portals. We make use of archival data and bibliographic information obtained from the NASA-IPAC Extragalactic Database (NED), data, and software facilities maintained by the Space Science Data Center (SSDC) from the Italian Space Agency. We thank the anonymous Referee for all comments, which helped to improve the discussion and presentation of the paper.

\bibliographystyle{mnras}
\bibliography{2bigb-main}
\newpage

% \section{Appendix \& Online Material}

%Here we show Table \ref{table:2BIGB-10yrs-slim}, which lists the results for the 11 years \& 0.5-500\,GeV broadband analysis for the 1160 3HSP sources detected down to TS\,=\,9 with {\it Fermi}-LAT. The sources are modeled with a power-law (eq. \ref{eq:powerlaw}) and the fitting parameters are the Normalization (N$_0$) and the photon spectral index ($\Gamma$) which are calculated at the pivot energy (E$_0$). Table \ref{table:2BIGB-SED} list the $\gamma$-ray SED data-point for the 1160 2BIGB sources, organized according to the template format needed to be uploaded at the online tools from ASI SED Builder \url{https://tools.ssdc.asi.it/SED/}. Both tables are available in text format (.ascii) at the Author's GitHub \url{https://github.com/BrunoArsioli} with instructions for upload to the ASI SED Tool service. In particular, Table \ref{table:2BIGB-10yrs-slim} is described with extra information from 3HSP and 4FGL catalogues.  

\newpage 
\input{2bigb-table.tex}

% Don't change these lines
\bsp	% typesetting comment
\label{lastpage}
\end{document}